\documentclass[traditabstract]{aa} 
\usepackage{graphicx}
\usepackage{txfonts}
\usepackage{aalongtable}

\begin{document}

\title{Star-forming fractions and galaxy evolution with redshift in rich X-ray-selected galaxy clusters}
\author{Julie B. Nantais\inst{1}, Alessandro Rettura\inst{2}, Chris Lidman\inst{3}, Ricardo Demarco\inst{1}, Raphael Gobat\inst{4}, Piero Rosati\inst{5}, \& M. James Jee\inst{6}}

\institute{Departamento de Astronom\'ia, Universidad de Concepci\'on, Casilla 160-C, Concepcion, Chile
 \and
    Department of Astronomy, California Institute of Technology, MS 105-24, Pasadena, CA 91125, USA
 \and
    Australian Astronomical Observatory, PO Box 296, Epping NSW 1710, Australia
 \and
    Laboratoire AIM-Paris-Saclay, CEA/DSM-CNRS-Universit\'e Paris Diderot, Irfu/Service d'Astrophysique, CEA Saclay, Orme des Merisiers, 91191 Gif-sur-Yvette, France
 \and
    ESO, Karl-Schwarzchild Strasse 2, 85748, Garching, Germany
 \and
    Department of Physics, University of California, Davis, One Shields Avenue, Davis, CA 95616, USA
}

   \date{Received ???; accepted ???}

  \abstract{We have compared stacked spectra of galaxies, grouped by environment and stellar mass, among 58 members of the redshift z = 1.24 galaxy cluster RDCS J1252.9-2927 (J1252.9) and 134 galaxies in the z = 0.84 cluster RX J0152.7-1357 (J0152.7).  These two clusters are excellent laboratories to study how galaxies evolve from star-forming to passive at z $\sim$ 1.  We measured spectral indices and star-forming fractions for our density- and mass-based stacked spectra.  The star-forming fraction among low-mass galaxies ($<$ 7 $\times$ 10$^{10}$ M$_{\odot}$) is higher in J1252.9 than in J0152.7, at about 4$\sigma$ significance.  Thus star formation is being quenched between z = 1.24 and z = 0.84 for a substantial fraction of low-mass galaxies.  Star-forming fractions were also found to be higher in J1252.9 in all environments, including the core.  Passive galaxies in J1252.9 have systematically lower D$_n$4000 values than in J0152.7 in all density and mass groups, consistent with passive evolution at modestly super-solar metallicities.}


 \authorrunning{Nantais et al.}
 \titlerunning{Galaxy evolution in z$\sim$1 clusters}

   \maketitle

\section{Introduction}
The cores of rich galaxy clusters have long been observed to contain mostly passive elliptical and lenticular galaxies, whereas less dense regions contain mostly star-forming spiral and irregular galaxies (e.g., Dressler \cite{dre80}).  The former generally make up most of what is referred to as the red sequence, so called because of the linear arrangement these galaxies have in a color-magnitude diagram.  The latter are typically found in a region of the color-magnitude diagram know as the blue cloud.  This trend, called the morphology-density relation, has been observed up to redshift z = 1.46 (Hilton et al.~\cite{hil09}).  The morphology-density relation is found in both mass-selected (Holden et al.~\cite{hol07}, van der Wel et al.~\cite{van07}) and luminosity-selected (Postman et al.~\cite{pos05}, Smith et al.~\cite{smi05}, Mei et al.~\cite{mei12}) samples.  

The earliest known examples of massive red-sequence galaxies in galaxy protoclusters are found between redshifts 2 and 3 (Kodama et al.~\cite{kod07}, Zirm et al.~\cite{zir08}). Recent evidence shows that star-forming galaxies are commonly found in cluster cores at redshifts between 1.5 and 2 (Papovich et al.~\cite{pap10}, Tanaka et al.~\cite{tan10}, Gobat et al.~\cite{gob11}, Fassbender et al.~\cite{fas11}), indicating that the massive cluster core red sequence grows primarily at z $>$ 1.5.  Evolution of cluster galaxies from the blue cloud to the red sequence continues outside the core at lower redshifts.  Passive red galaxies are found in all environments (core and outskirts) at z $<$ 1.5, and many star-forming galaxies are still being accreted in cluster outskirts even at low redshifts.  Galaxies accreted at later times must be transformed from blue late types to red early types, albeit possibly by different methods than the cluster core galaxies at z $>$ 1.5.  

High-density environmental effects such as ram-pressure stripping (Gunn \& Gott \cite{gun72}), harassment (Moore et al.~\cite{mor96}), and strangulation (Balogh \& Morris \cite{bal00a}, Balogh, Navarro, \& Morris \cite{bal00b}) can disturb and/or remove the galaxies' cold gas.  These effects may have already begun to occur for galaxies not found in cluster cores.  Wetzel et al.~(\cite{wet13a}) suggest that many galaxies found outside the virial radius of cluster halos are backsplash or ejected satellite galaxies, i.e., galaxies that have crossed the cluster core at least once and later returned to the outskirts.  The single core crossings are found to be sufficient to explain the enhanced fractions of passive and near-passive galaxies in the outskirts of clusters relative to the field (Wetzel et al.~\cite{wet13a}).

Environment-related quenching processes such as strangulation can work in small groups as well as in massive clusters.  A study of galaxy groups at 0.85 $<$ z $<$ 1 by Balogh et al.~(\cite{bal11}) found that many group-member galaxies had already developed color and morphology intermediate between the blue cloud and the red sequence.  Their field counterparts, on the other hand, had more typical blue-cloud features.  A study by Lemaux et al.~(\cite{lem12}) found a deficit of faint red galaxies and an abundance of intermediate-mass green galaxies associated with a z = 0.9 supercluster.  Both of these results suggest a possible role of pre-processing in cluster galaxy evolution, in which galaxies begin to form fewer stars in small group environments before falling into a cluster.

Environment, however, is not the only factor determining galaxy evolution in cluster outskirts: self-quenching in massive accreted cluster galaxies can also play an important role in the cessation of star formation.  Muzzin et al.~(\cite{muz12}) performed a comprehensive study combining the spectra of ten infrared-selected galaxy clusters at redshifts between 0.8 and 1.4.  They compared galaxies in various cluster-centric radius and stellar mass bins with each other and with field galaxies.  The 4000 {\AA} break and specific star-formation rates were found to vary notably with both stellar mass and environment for the full samples.  Star-forming fractions also showed strong environmental variation.  However, when passive and star-forming galaxies were considered separately, variations in the 4000 {\AA} break were only seen with stellar mass.  The specific star-formation rates for star-forming galaxies also varied only with stellar mass and not with cluster-centric radius.  These findings suggest that environment primarily affects star-forming fraction, i.e., whether galaxies have been quenched or not at z $\sim$ 1, rather than the epoch of initial galaxy formation or the star formation rate.  Before or after quenching, the galaxies' properties are primarily a function of stellar mass.  Since the galaxies were stacked across different redshifts, the Muzzin et al.~(\cite{muz12}) study could not determine how the environmental influences on galaxy evolution changed with time.

Trends in cluster galaxy evolution at 0.4 $<$ z $<$ 1 have been observed in optically-selected clusters using the Red-Sequence Cluster Survey (Hsieh et al.~\cite{hsi05}).  Galaxy clusters at these redshifts typically had mature populations of bright red galaxies in their cores, with little increase in the fraction of blue galaxies in cluster cores at higher redshifts (Loh et al.~\cite{loh08}).  However, galaxy clusters had more faint red galaxies at lower redshifts and more faint blue galaxies at higher redshifts within 0.5 R$_{200}$(Gilbank et al.~\cite{gil08}).  The clusters as a whole also had a few more bright blue galaxies at high redshifts with colors similar to modern-day spiral galaxies (Loh et al.~\cite{loh08}).  These findings suggest that at intermediate redshifts, migration from the star-forming blue cloud to the passive red sequence occurs among faint galaxies outside the innermost regions of clusters.

To improve our understanding of the history of transformation of cluster galaxies, we need to extend similar studies to higher redshifts, and study these trends in clusters selected using different methods.  Two of the richest and best-studied X-ray selected galaxy clusters at z $\sim$ 1 are RX J0152.7-1357 (J0152.7) at z = 0.84 and RDCS J1252.9-2927 (J1252.9) at z = 1.24, separated by about 1.5 Gyr in cosmic time.  The galaxy cluster J0152.7 is a dynamically young (Demarco et al.~\cite{dem05}, Girardi et al.~\cite{gir05}), X-ray bright cluster (Rosati et al.~\cite{ros98}, Della Ceca et al.~\cite{del00}).  It consists of two merging cluster cores, two rich groups (Demarco et al.~\cite{dem10}), and the low-density outskirts, with 134 spectroscopically confirmed members (Demarco et al.~\cite{dem05}, \cite{dem10}).  About a third of these spectroscopically confirmed members are still forming stars, and about half of all confirmed members are located in the low-density outskirts of the cluster.

A wide-field study by Patel et al.~(\cite{pat09}) with low-dispersion prism spectroscopy of J0152.7 and its large-scale structure found an excess of red galaxies in areas with intermediate projected galaxy densities as compared with the lowest-density regions.  Demarco et al.~(\cite{dem10}) estimated younger ages and more recent quenching in the faintest, bluest red-sequence galaxies of J0152.7 based on spectrophotometric star-formation histories.  Nantais et al.~(\cite{nan13}) performed a detailed morphological analysis on the confirmed members of J0152.7 using a variation on the methods of Neichel et al.~(\cite{nei08}) and Delgado-Serrano et al.~(\cite{del10}).  This analysis method, sensitive to features correlated with kinematic differences in galaxies as found by Neichel et al.~(\cite{nei08}), revealed that 38\% of cluster galaxies overall had peculiar features or were compact.  This fraction changed especially drastically between the lowest-density and intermediate-density regions, dropping from 53\% to 23\%.  These studies indicate recent and ongoing evolution from the blue cloud to the red sequence, and from irregular or spiral to early-type morphology, among galaxies in this cluster.

The galaxy cluster J1252.9 is also an X-ray bright cluster (Rosati et al.~\cite{ros04}), with an elongated structure (Tanaka et al.~\cite{tan07}).  It has 38 members spectroscopically confirmed in the literature, and evidence suggests that it has also undergone relatively recent merger activity (Demarco et al.~\cite{dem07}).  Spectrophotometric star-formation histories for the J1252.9 galaxies suggest that the red sequence in this cluster formed around z = 2  (Gobat et al.~\cite{gob08}).  Rettura et al.~(\cite{ret10}, \cite{ret11}), also working with star-formation histories, found that the timescale of the star-formation history, including the last star-forming episode, depended on environment.  The initial timing of galaxy formation, however, depended on stellar mass.  The environment of J1252.9 has a lower fraction of faint red galaxies in the cluster outskirts than in the cluster core (Tanaka et al.~\cite{tan07}).  In a spectroscopic follow-up, red, dusty star-forming galaxies were found in some of the surrounding groups, suspected to be provoked by mergers and low-speed tidal encounters (Tanaka et al.~\cite{tan09}).  It is thus clear that J0152.7 and J1252.9 both have plenty of ongoing galaxy evolution and are excellent astrophysical laboratories to study galaxy evolution in dense environments.

The clusters J0152.7 and J1252.9 are also excellent to compare to one another because they are similar in mass.  The merging cluster cores of J0152.7 are estimated to have dynamical masses of 2.5$\pm$0.9 and 1.1$\pm$0.4 $\times$ 10$^{14}$ M$_{\odot}$ (Demarco et al.~\cite{dem05}).  J1252.9 is of at least comparable mass to J0152.7 if not slightly more massive.  The dynamical mass estimated using the same method as Demarco et al.~(\cite{dem07}) for all members (including those confirmed in this paper) within the 1.61 Mpc virial radius is 5.2$\pm$1.0 $\times$ 10$^{14}$ M$_{\odot}$.  Thus, we can be confident that any difference in star-forming fractions among low-mass galaxies in J1252.9 and J0152.7 are not a result of a lower halo mass for J1252.9.  In fact, the halo mass of J1252.9 can be expected to grow in 1.5 Gyr (Fakhouri et al.~\cite{fak10}), making the cluster notably more massive than J0152.7 for its epoch.

In this paper, we investigate the properties of cluster galaxies as a function of their environment, stellar mass, and cosmic time, by spectroscopically comparing environment- and stellar-mass-grouped galaxy samples in J1252.9 and J0152.7.  In Sect.~2, we present the new spectroscopic data for J1252.9 and the confirmation of twenty new members of this galaxy cluster.  In Sect.~3, we compare the galaxy populations in J1252.9 and J0152.7 according to environment and stellar mass.  We define a new index that measures the strength of the higher-order Balmer line H6 with low contamination from the CN band in old stellar populations.  In Sect.~4 we discuss the results of our comparison, including differences in star-forming fractions and spectral indices among various sub-populations.  In Sect.~5 we interpret our results.  Finally, in Sect.~6 we summarize the conclusions of our study.  Throughout this paper, we assume a $\Lambda$CDM cosmology with $\Omega_M$ $=$ 0.3, $\Omega_{\Lambda}$ $=$ 0.7, and $H_0$ $=$ 70 km s$^{-1}$ Mpc$^{-1}$.

\section{New spectroscopic data}

We have obtained spectroscopy of 120 objects in and around J1252.9, including 107 objects with previously unconfirmed redshifts, with the Focal Reducer and Low Dispersion Spectrograph 2 (FORS2) on the European Southern Observatory's Very Large Telescope (VLT) in mask exchange unit (MXU) mode$^{1}$.  Nearly all candidates for spectroscopic observation had optical photometry in $B$, $V$, and $R$ with FORS2 (Demarco et al.~\cite{dem07}).  Most candidates were also observed in $i_{775}$ and $z_{850}$ with the {\it{Hubble}} Space Telescope Advanced Camera for Surveys (Blakeslee et al.~\cite{bla03}).  Many candidates also had near-infrared imaging in $J$ and $K_s$ with the VLT Infrared Spectrometer and Array Camera (ISAAC) (Lidman et al.~\cite{lid04}) and in the {\it{Spitzer}} Space Telescope 3.6$\mu$m and 4.5$\mu$m bands.  

\footnotetext[1]{The VLT observations were taken under Program ID 087.A-0581.}

Four MXU masks were observed between July 2011 and March 2012, all with the 300I grism and the OG 590 filter.  Total exposure times were 5 hours 15 minutes for Masks 1 and 3, 6 hours 15 minutes for Mask 2, and 6 hours 45 minutes for Mask 4.  Mask 1 was observed in January 2012, Mask 2 was observed in July 2011 and February 2012, and Masks 3 and 4 were observed in March 2012.  The instrumental full width at half-maximum (FWHM) resolution for this filter, grating, and grism combination is approximately 11 {\AA}, measured by fitting profiles to several unblended arc lamp lines.  Most of the Demarco et al.~(\cite{dem07}) spectra have about the same resolution as the new spectra, except for those observed in Moveable Slitlets (MOS) mode with the Focal Reducer and low dispersion Spectrograph 1 (FORS1), which have a resolution of approximately 15 {\AA}.   Masks were reduced using the same custom software used in Demarco et al.~(\cite{dem07}).  

The primary goal of our observations was to identify new members, and especially new red-sequence members, in the outskirts of the cluster.  After targeting as many new candidates within our color selections as possible, we targeted previously confirmed members in an attempt to obtain better spectra and/or estimate redshift uncertainties.  If no new candidates or old members were available, we then targeted X-ray sources and objects outside the color cuts.  Red candidates were chosen using a simple optical color cut, $V-i$ $>$ 1.8 and $i-z$ $>$ 0.8, in order to avoid limiting our search to the regions covered by the $K_s$-band imaging.  Blue candidates were selected using a variation of the optical color selection described in Demarco et al.~(\cite{dem07}) to search for star-forming galaxies at z $\sim$ 1.24: $0.4 < i-z < 0.85$, $0.2 < V-i < 1.2$, and $[2.4\times(i-z)-1.12] < V-i < [7.0\times(i-z)-2.3]$. Previously unconfirmed red candidates were given first priority to observe, and unconfirmed blue candidates were given second priority.  

Figure 1 shows the $V-i$ and $i-z$ color-color diagram for J1252.9 spectroscopic targets from this paper and Demarco et al.~(\cite{dem07}) with the red and blue selection boxes.  All of the confirmed passive members lie within the red selection box.  Confirmed star-forming members are more variable in their color ranges, though roughly half of the star-forming members are in the blue selection box.

\begin{figure}
\centering
\includegraphics[width=8.5cm]{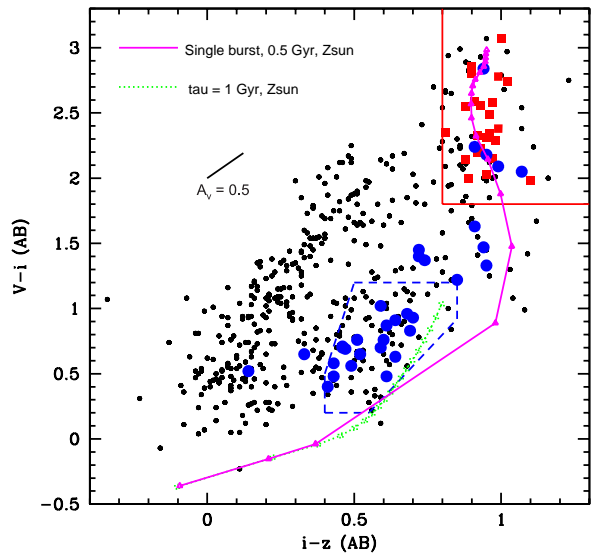}
\caption{$V-i$ vs. $i-z$ color-color diagram of spectroscopically-targeted J1252.9 objects (from this paper and Demarco et al.~\cite{dem07}) and the color cuts used to select candidates for spectroscopic observation in this paper.  The solid red box represents the color range for choosing red candidates, and the dashed blue polygon represents the color range for selection of blue candidates. Black dots represent non-members and unconfirmed objects.  Red squares are confirmed passive members, and blue circles are confirmed star-forming members.  Also shown are two Bruzual \& Charlot (\cite{bc03}) galaxy evolution tracks at solar metallicity: a 0.5 Gyr single burst and a $\tau$ = 1 Gyr exponentially-declining star-formation model, at present-day ages ranging from 8.5 to 13 Gyr (formation redshift 1.22 to 9.84).  The short black line corresponds to a reddening vector for A$_v$ = 0.5.}
\label{FigColCut}
\end{figure}

\begin{figure*}
\centering
\includegraphics[width=18cm]{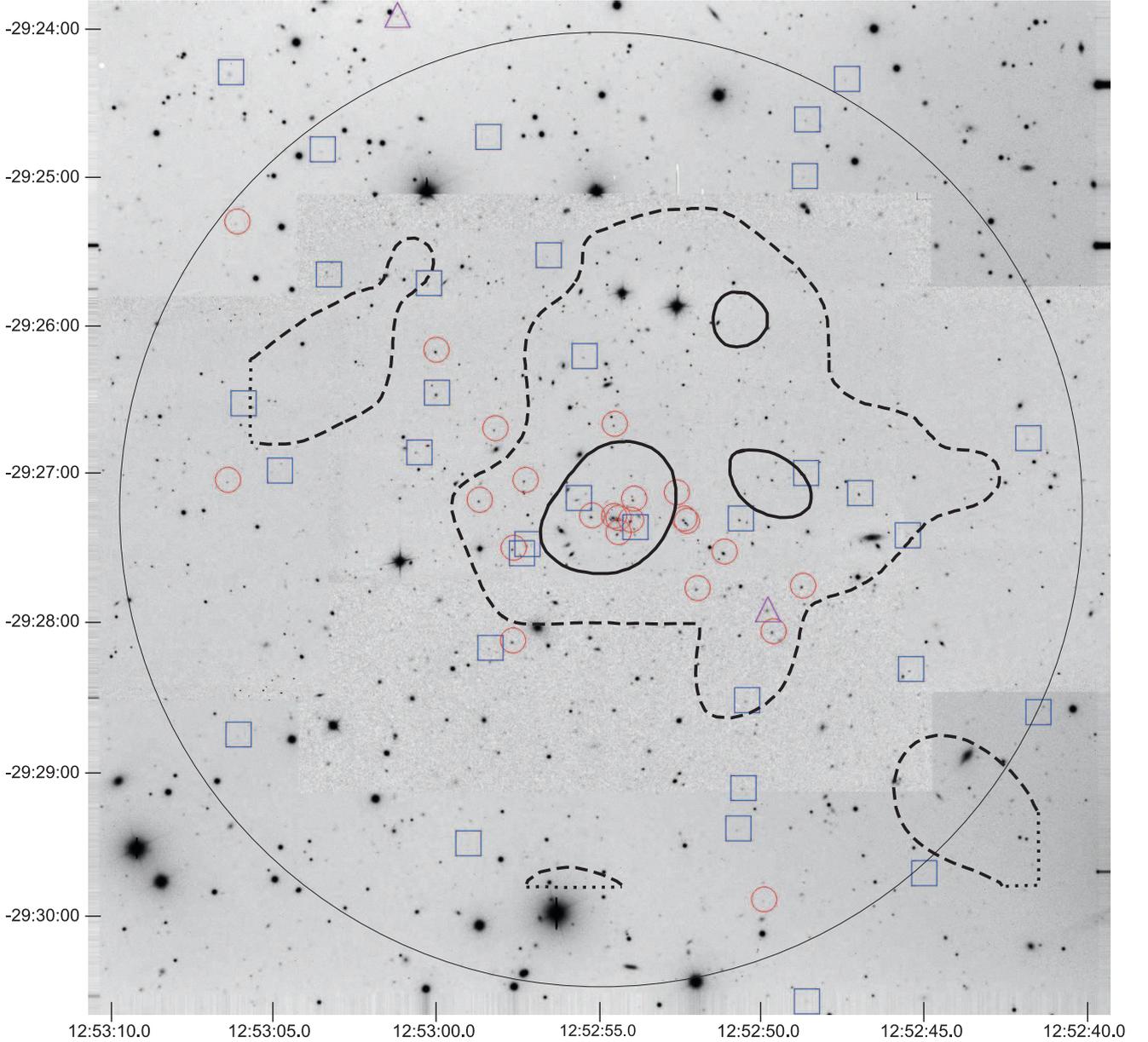}
\caption{Distribution of 58 confirmed members of the galaxy cluster J1252.9 shown on a VLT FORS2 I-band + ISAAC K-band mosaic image.  North is up and east is to the left.  Blue squares are star-forming galaxies, red circles are passive galaxies, and purple triangles are X-ray sources.  The image measures approximately 6.83 $\times$ 6.83 arcmin, or 3.41 $\times$ 3.41 proper Mpc at z = 1.24.  The thin black circle represents the estimated virial radius of 1.61 Mpc from Demarco et al.~(\cite{dem07}).  Thick dashed and solid black lines represent boundaries of the intermediate and high density regions, respectively, as defined in Sect.~3.1.  The dotted lines represent the edges of the Jee et al.~\cite{jee11} weak-lensing map as encountered within an intermediate-density region.}
\label{FigLoc}
\end{figure*}

\begin{figure*}
\centering
\includegraphics[width=18cm]{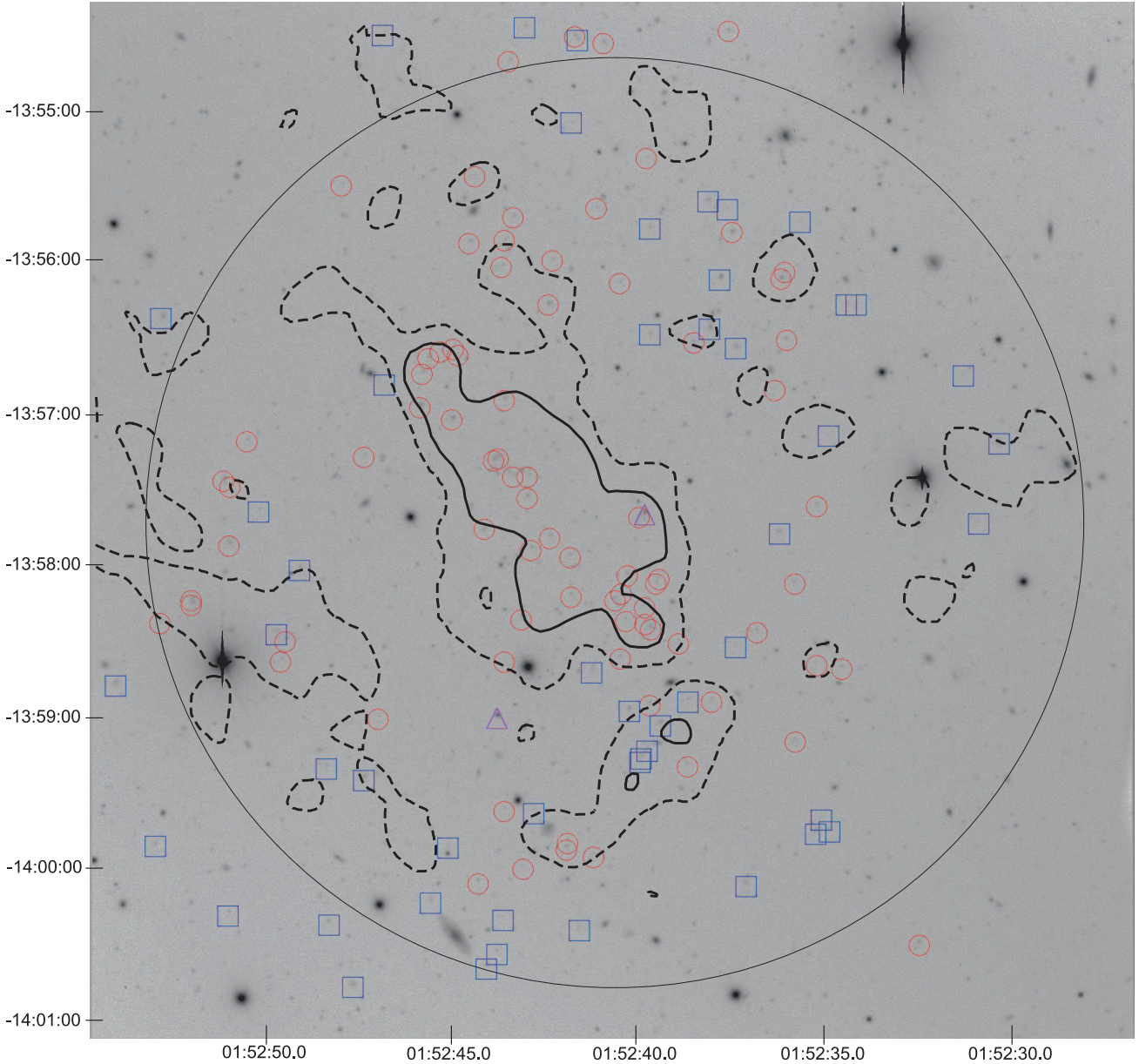}
\caption{Distribution of most of the 134 confirmed members of galaxy cluster J0152.7 shown on a VLT FORS1 R-band image.  North is up and east is to the left.  Blue squares are star-forming galaxies, red circles are passive galaxies, and purple triangles are X-ray sources.  The image measures approximately 6.8 $\times$ 6.8 arcmin, or 3.12 $\times$ 3.12 proper Mpc at z = 0.84.  The black circle represents the estimated virial radius of 1.4 Mpc from Demarco et al.~(\cite{dem05}).  Thick dashed and solid black lines represent the boundaries of intermediate- and high-density regions, respectively, as defined in Sect.~3.1.}
\label{FigLoc0152}
\end{figure*}

\begin{table*}
\caption{J1252.9 Cluster Members}
\label{tab1}
\centering
\begin{tabular}{lccccccl}
\hline\hline  
ID & RA & DEC & z & Em. & Old ID & Photometry & Spectral Features\\ 
 & (hh:mm:ss) & (dd:mm:ss) & & & & & \\
\hline  
XrayS86 & 12:53:01.197 & -29:23:54.00 & 1.2337 & 2 & N/A & none & MgII em., [OII], MgI, CaII (H+K), D4000 \\
423 & 12:52:41.507 & -29:28:36.34 & 1.2417 & 1 & N/A & $BVRiz$ & [OII] \\
1517 & 12:52:45.004 & -29:29:41.61 & 1.2404 & 1 & N/A & $BVRiz$ & [OII] \\
1677 & 12:52:45.411 & -29:28:18.98 & 1.2382 & 0.5 & N/A & $BVRizJK_s$(3.6$\mu$m)(4.5$\mu$m) & [OII]?, MgI, CaII (H+K), D4000 \\
1720 & 12:52:45.522 & -29:27:24.75 & 1.2333 & 0.5 & N/A & $BVRizJK_s$3.6$\mu$m4.5$\mu$m & [OII]?, CaII (H+K), D4000, H$\delta$ \\
2238 & 12:52:46.980 & -29:27:07.70 & 1.2379 & 1 & 345 & $BVRizJK_s$3.6$\mu$m4.5$\mu$m & H10, MgI, [OII], D4000 \\
2358 & 12:52:47.386 & -29:24:20.03 & 1.2389 & 1 & N/A & $BVRiz$ & [OII], H9, Mg 1, H6 \\
2758 & 12:52:48.608 & -29:24:36.35 & 1.2394 & 1 & N/A & $BVRiz$ & FeI, [OII], H8, MgI, H6, CaII (H+K), D4000, H$\delta$ \\
2759 & 12:52:48.640 & -29:26:59.70 & 1.2371 & 1 & 366 & $BVRizJK_s$3.6$\mu$m4.5$\mu$m & [OII] \\
2779 & 12:52:48.680 & -29:24:59.30 & 1.2380 & 1 & 6306 & $BVRiz$3.6$\mu$m4.5$\mu$m & [OII], H9, H8, MgI, H6, CaII (H+K), H$\delta$ \\
2797 & 12:52:48.740 & -29:27:45.30 & 1.2353 & 0 & 206 & $BVRizJK_s$3.6$\mu$m4.5$\mu$m & H8, MgI, H6, CaII (H+K), D4000 \\
3114 & 12:52:49.650 & -29:28:03.70 & 1.2382 & 0 & 149 & $VRizJK_s$3.6$\mu$m4.5$\mu$m & MgI, CaII (H+K), D4000, H$\delta$ \\
3132 & 12:52:49.820 & -29:27:54.90 & 1.2382 & 1 & 174 & $BVRizJK_s$3.6$\mu$m4.5$\mu$m & [OII] \\
3214 & 12:52:49.930 & -29:29:52.30 & 1.2416 & 0 & 7005 & $BVRiz$3.6$\mu$m4.5$\mu$m & HeI, H10, H8, MgI, H6, CaII (H+K), D4000, H$\delta$ \\
3424 & 12:52:56.556 & -29:25:31.35 & 1.2394 & 1 & N/A & $BVRizJK_s$3.6$\mu$m4.5$\mu$m & [OII], H6 \\
3767 & 12:52:55.440 & -29:26:12.20 & 1.2402 & 1 & 515 & $BVRizJK_s$3.6$\mu$m4.5$\mu$m & HeI, H9, CaII (H+K), [OII] \\
3860 & 12:52:55.220 & -29:27:16.70 & 1.2455 & 0 & 313 & $VRizJK_s$3.6$\mu$m4.5$\mu$m & FeI, H10, MgI, CaII (H+K), D4000, CaI \\
4159 & 12:52:54.540 & -29:27:17.10 & 1.2378 & 0 & 289 & $BVRizJK_s$(3.6$\mu$m)(4.5$\mu$m) & H9, D4000, CaI \\
4199 & 12:52:54.410 & -29:27:17.70 & 1.2343 & 0 & 291 & $BVRizJK_s$(3.6$\mu$m)(4.5$\mu$m) & FeI, H9, H8, CaII (H+K), D4000, H$\delta$ \\
4235 & 12:52:54.520 & -29:26:39.80 & 1.2306 & 1 & 442 & $BVRizJK_s$3.6$\mu$m4.5$\mu$m & HeI, MgI, H6, CaII (H+K), D4000 \\
4325 & 12:52:54.420 & -29:27:23.70 & 1.2472 & 0 & 282 & $BVRizJK_s$3.6$\mu$m4.5$\mu$m & CaII (H+K), D4000, CaI \\
4479 & 12:52:54.020 & -29:27:18.50 & 1.2384 & 0 & 304 & $BVRizJK_s$3.6$\mu$m4.5$\mu$m & MgI, H6, CaII (H+K), D4000, CaI \\
4503 & 12:52:53.930 & -29:27:09.90 & 1.2312 & 0 & 338 & $BVRizJK_s$3.6$\mu$m4.5$\mu$m & FeI, HeI, H8, MgI, CaII (H+K), D4000, H$\delta$, CaI \\
4525 & 12:52:55.630 & -29:27:09.70 & 1.2274 & 1 & 339 & $BVRizJK_s$3.6$\mu$m4.5$\mu$m & [OII], H9, H8 \\
4553 & 12:52:53.896 & -29:27:21.44 & 1.2361 & 0.5 & N/A & $BVRizJK_s$ & [OII]?, MgI, H6, CaII (H+K), D4000, H$\delta$ \\
5137 & 12:52:52.626 & -29:27:07.43 & 1.2446 & 0 & N/A & $BVRizJK_s$3.6$\mu$m4.5$\mu$m & FeI, HeI: MgI, H6, CaII (H+K), D4000, H$\delta$ \\
5221 & 12:52:52.390 & -29:27:18.00 & 1.2342 & 0 & 310 & $BVRizJK_s$(3.6$\mu$m)(4.5$\mu$m) & FeI, HeI, H8, MgI, H6, CaII (H+K), D4000, CaI \\
5222 & 12:52:52.310 & -29:27:19.10 & 1.2455 & 0 & 294 & $BVRizJK_s$(3.6$\mu$m)(4.5$\mu$m) & MgI, CaII (H+K), D4000, CaI \\
5397 & 12:52:51.980 & -29:27:46.20 & 1.2318 & 0 & 205 & $BVRizJK_s$3.6$\mu$m4.5$\mu$m & MgI, CaII (H+K), D4000, CaI \\
5801 & 12:52:51.150 & -29:27:31.40 & 1.2351 & 0 & 247 & $BVRizJK_s$3.6$\mu$m4.5$\mu$m & H8, H6, CaII (H+K), D4000, H$\delta$, CaI \\
5997 & 12:52:50.733 & -29:29:23.64 & 1.2418 & 0.5 & N/A & $BVRiz$3.6$\mu$m4.5$\mu$m & [OII]?, H9, H8, MgI, H6, CaII (H+K), D4000, H$\delta$ \\
6052 & 12:52:50.573 & -29:29:07.10 & 1.2417 & 1 & N/A & $BVRizJK_s$3.6$\mu$m4.5$\mu$m  & [OII], H9: H8, MgI, H6 \\
6064 & 12:52:50.660 & -29:27:18.10 & 1.2312 & 1 & 309 & $BVRizJK_s$3.6$\mu$m4.5$\mu$m  & [OII], H6, CaII (H+K), H$\delta$, CaI \\
6115 & 12:52:50.458 & -29:28:31.57 & 1.2401 & 0.5 & N/A & $BVRizJK_s$3.6$\mu$m4.5$\mu$m  & [OII]?, H10, MgI, H6, CaII (H+K), D4000, H$\delta$ \\
6383 & 12:52:57.212 & -29:27:28.24 & 1.2395 & 1 & N/A & $BVRizJK_s$3.6$\mu$m4.5$\mu$m  & [OII], H10, H9, H8, MgI, H6 \\
6392 & 12:52:57.270 & -29:27:02.30 & 1.2295 & 0 & 367 & $BVRizJK_s$(3.6$\mu$m)(4.5$\mu$m) & MgI, CaII (H+K), D4000 \\
6592 & 12:53:06.307 & -29:24:17.19 & 1.2358 & 1 & N/A & $BVRiz$ & [OII], H10, H8, MgI, H6, CaII (H), H$\delta$ \\
6673 & 12:53:06.128 & -29:25:17.80 & 1.2359 & 0 & N/A & $BVRiz$ & H10, H8, MgI, H6, CaII (H+K), D4000, H$\delta$ \\
6688 & 12:53:06.087 & -29:28:45.42 & 1.2338 & 1 & N/A & $BVRiz$ & [OII], H10, H8, H6\\
7546 & 12:53:03.500 & -29:24:48.60 & 1.2408 & 1 & 6301 & $BVRiz$(3.6$\mu$m) & [OII] \\
7581 & 12:53:03.300 & -29:25:39.30 & 1.2327 & 1 & 2065 & $BVRizJK_s$(3.6$\mu$m)(4.5$\mu$m) & [OII] \\
8193 & 12:53:00.230 & -29:25:42.70 & 1.2432 & 1 & 619 & $BVRizJK_s$3.6$\mu$m4.5$\mu$m  & [OII], CaI \\
8529 & 12:53:00.530 & -29:26:51.20 & 1.2459 & 1 & 3159 & $BVRizJK_s$(3.6$\mu$m)(4.5$\mu$m) & [OII], H10, H9, H8 \\
8686 & 12:52:59.980 & -29:26:27.20 & 1.2400 & 1 & 445 & $BVRizJK_s$3.6$\mu$m4.5$\mu$m  & [OII], H10, MgI, H6, CaII (H+K) \\
8709 & 12:53:00.000 & -29:26:09.80 & 1.2354 & 0 & 407 & $BVRizJK_s$3.6$\mu$m4.5$\mu$m  & H8, MgI, CaII (H+K), D4000, H$\delta$ \\
9067 & 12:52:59.020 & -29:29:29.48 & 1.2205 & 1 & N/A & $BVRizJK_s$3.6$\mu$m4.5$\mu$m  & [OII] \\
9197 & 12:52:58.399 & -29:24:43.44 & 1.2363 & 1 & N/A & $BVRiz$ & [OII], MgI, H6, CaII (H+K) \\
9204 & 12:52:58.358 & -29:28:10.34 & 1.2385 & 1 & N/A & $BVRizJK_s$(3.6$\mu$m)(4.5$\mu$m) & [OII], MgI, H6 \\
9227 & 12:52:58.670 & -29:27:10.40 & 1.2297 & 0 & 330 & $BVRizJK_s$3.6$\mu$m4.5$\mu$m  & HeI, H9, H8, MgI, H6, CaII (H+K), D4000, H$\delta$ \\
9354 & 12:52:58.190 & -29:26:41.60 & 1.2416 & 0 & 432 & $BVRizJK_s$3.6$\mu$m4.5$\mu$m  & H10, MgI, H6, CaII (H+K), D4000 \\
9465 & 12:52:57.640 & -29:27:29.80 & 1.2358 & 0 & 265 & $BVRizJK_s$3.6$\mu$m4.5$\mu$m  & H8, CaII (K), D4000 \\
9466 & 12:52:57.650 & -29:28:07.50 & 1.2475 & 0 & 137 & $BVRizJK_s$3.6$\mu$m4.5$\mu$m  & MgI, CaII (H+K), D4000, H$\delta$ \\
9470 & 12:52:57.390 & -29:27:32.10 & 1.2322 & 1 & 248 & $BVRizJK_s$3.6$\mu$m4.5$\mu$m  & [OII], CaII (H), CaI \\
30145 & 12:53:05.930 & -29:26:31.40 & 1.2273 & 1 & 726 & $BVRizJK_s$3.6$\mu$m4.5$\mu$m  & [OII] \\
30337 & 12:53:04.810 & -29:26:58.60 & 1.2438 & 1 & 9000 & $BVRizJK_s$3.6$\mu$m4.5$\mu$m  & [OII] \\
30917 & 12:53:06.410 & -29:27:02.40 & 1.2373 & 0 & 370 & $VRizJK_s$3.6$\mu$m4.5$\mu$m  & FeI, H10, MgI, CaII (H+K), D4000 \\
40950 & 12:52:41.810 & -29:26:45.50 & 1.2338 & 1 & 419 & $BVRizJK_s$3.6$\mu$m4.5$\mu$m  & [OII] \\
7001\_old & 12:52:48.610 & -29:30:33.50 & 1.2368 & 1 & 7001 & none & FeI, HeI, [OII], H9, H8, MgI, H6, CaII (H+K), H$\delta$ \\
\hline  
\end{tabular}
\tablefoot{Emission line flags (Column 5) are as follows: 0 = no emission, 0.5 = marginal [OII] emission, 1 = clear [OII] emission, 2 = clear active galactic nucleus/broad line emission.  The flags are based on those used in Demarco et al.~(\cite{dem05}) and Demarco et al.~(\cite{dem07}), with the ``0.5'' flag added to denote the potential confusion of [OII] with residuals from the 8348 {\AA} telluric line in J1252.9.  The old ID numbers in Column 6 are from Demarco et al.~(\cite{dem07}).  Photometric bands in parentheses represent only unreliable or blended photometry in that band.}
\end{table*}

Redshifts were measured by fitting templates with the XCSAO (cross-correlation, Smithsonian Astrophysical Observatory) task in the Image Reduction and Analysis Facility (IRAF$^{2}$)  RVSAO (radial velocity, Smithsonian Astrophysical Observatory) package when possible.  The initial guess for redshift was based on emission lines, absorption lines, and/or continuum shape in the two-dimensional (2-D) and one-dimensional (1-D) reduced, wavelength-calibrated spectra.  Galaxies with no measurable emission lines were confirmed at their redshift if they had R $>$ 3.5 for the best-fitting template  (Tonry \& Davis \cite{ton79}).  Above this R-value cutoff, the prominent absorption features and breaks were clear and well aligned with their predicted positions according to the cross-correlation redshift.  In most cases, the fit was performed in the wavelength range 6000-9500 {\AA}, where the flux calibration was most reliable and the signal-to-noise ratio (S/N) was the highest.  If emission, absorption, or continuum shape features were observable outside this wavelength range, the wavelength range for redshift determination was expanded to where these features were observed.  

If only a single emission line was found, it was presumed to be [OII], since the wavelength coverage of FORS2 with the 300I grism is broad enough to rule out other possibilities such as [OIII] or H$\alpha$.  Ly$\alpha$ could also be ruled out based on the shape of the emission line and the surrounding continuum and absorption features.  For very faint [OII] emitters with little continuum or absorption lines, redshift was determined via a profile fit to the [OII] 3727 {\AA} line in the 1-D spectrum, whose location was pinpointed in the 2-D spectrum.  

\footnotetext[2]{IRAF is distributed by the National Optical Astronomy Observatory, which is operated by the Association of Universities for Research in Astronomy, Inc., under cooperative agreement with the National Science Foundation.}

Of the 107 unique, previously unconfirmed objects, we identified 15 as stars and confirmed redshifts for 47 galaxies.  Using the selection criteria of Demarco et al.~(\cite{dem07}) to identify cluster members, twenty of these 47 galaxies were members of J1252.9.  This increased the number of spectroscopically confirmed galaxies in J1252.9 from 38 to 58.  Two were passive, twelve were strong to moderate [OII] emitters, and one was an active galactic nucleus (AGN) with narrow [OII] and broad MgI 2800 {\AA} emission.  The remaining five spectra showed very faint [OII] superimposed on the 8348 {\AA} telluric line and moderate or strong Balmer absorption. In Tab.~1, we designate these marginal [OII] emitters with an emission flag of 0.5 to denote that the [OII] is very faint and may be confused with sky residuals.

The IDs, positions, redshifts,  emission-line flags, photometric coverage, and prominent spectral features of all fifty-eight J1252.9 member galaxies are listed in Tab.~1.  Only two confirmed members lie outside of the range of the optical photometry.  One is ID 7001 from Demarco et al.~(\cite{dem07}).  The other is the new AGN, XrayS86, from a catalog of X-ray point sources derived from Rosati et al.~(\cite{ros04}) and Martel et al.~(\cite{mar07}).

The estimated uncertainty in the redshifts given in Tab.~1 is about 0.0014.  We estimated this uncertainty by comparing our redshifts to those of Demarco et al.~(\cite{dem07}) for eight objects: five confirmed non-members and three confirmed members in common between the two spectroscopic campaigns.  Our uncertainty value is similar to the total redshift uncertainty of 0.0012 cited in Demarco et al.~(\cite{dem07}).

Figure 2 shows the locations of [OII]-emitting galaxies (blue squares), passive galaxies (red circles), and X-ray sources (purple triangles) in J1252.9 from this work and Demarco et al.~(\cite{dem07}).  Four members (7\% of the sample) are located outside the virial radius: old ID 7001, the new AGN XrayS86, ID 2358, and ID 6592, all emission-line galaxies.  Figure 3 shows the locations of the J0152.7 galaxies on a similar scale to that of J1252.9, with the same symbols for different types of galaxies. There are 16 members outside the virial radius of J0152.7 (12\% of the total sample), including two passive members beyond the angular coverage of Fig.~3.  

Our spectroscopic campaign confirmed six new members toward the southwest of the cluster core, which previously had only two confirmed members.  We also confirmed five new members toward the northeast.  The large-scale structure of J1252.9 as traced by the 58 confirmed members appears to extend from northeast to southwest in an elongated pattern, as found in the photometric study of Tanaka et al.~(\cite{tan07}).  Demarco et al.~(\cite{dem07}), with the original 38 members, had observed primarily an east-west extension.  The central regions contain mostly passive galaxies, and the outskirts are dominated by star-forming galaxies.  There are also two members which are X-ray sources.  One is the previously-identified member 174 (ID 3132 in Table 1), slightly to the southwest of the cluster core.  Since it spectroscopically looks like an ordinary [OII] emitter, it has an emission flag of 1 in Tab.~1.  The other is the new member XrayS86, far to the north-northeast, which has faint broad MgII 2800 {\AA} emission, narrow [OII] emission, and absorption lines typical of passive galaxies.

\section{Analysis}
In a similar fashion to Demarco et al.~(\cite{dem10}) for J0152.7, we performed weighted stacks of subsamples of the galaxies from both clusters.  The galaxies from each cluster were divided into three density samples and two stellar-mass samples for all galaxies combined (star-forming and passive).  These density and stellar-mass samples were defined again separately for passive-only subsamples and for star-forming-only subsamples.  The definitions of subsamples and the motivations behind the choice of stacking weights are described below.

\subsection{Density and stellar mass subsample definitions}

\begin{table}
\caption{J1252.9 Stellar Masses}
\label{tab2}
\centering
\begin{tabular}{lccl}
\hline\hline 
ID & Stellar Mass & $\sigma$ & Bands \\ 
 &  & & \\
\hline
423 & 2 $\times$ 10$^{10}$ & 2 $\times$ 10$^{10}$ & $BVRiz$ \\
1517 & 6 $\times$ 10$^{09}$ & 6 $\times$ 10$^{09}$ & $BVRiz$ \\
1677 & 1 $\times$ 10$^{11}$ & 3 $\times$ 10$^{10}$ & $BVRizJK_s$ \\
1720 & 2 $\times$ 10$^{10}$ & 5 $\times$ 10$^{09}$ & $BVRizJK_s$3.6$\mu$m4.5$\mu$m \\
2238 & 1 $\times$ 10$^{11}$ & 1 $\times$ 10$^{10}$ & $BVRizJK_s$3.6$\mu$m4.5$\mu$m \\
2358 & 3 $\times$ 10$^{10}$ & 3 $\times$ 10$^{10}$ & $BVRiz$ \\
2758 & 6 $\times$ 10$^{10}$ & 2 $\times$ 10$^{11}$ & $BVRiz$ \\
2759 & 1 $\times$ 10$^{10}$ & 1 $\times$ 10$^{09}$ & $BVRizJK_s$3.6$\mu$m4.5$\mu$m \\
2779 & 4 $\times$ 10$^{10}$ & 2 $\times$ 10$^{10}$ & $BVRiz$3.6$\mu$m4.5$\mu$m \\
2797 & 1 $\times$ 10$^{11}$ & 2 $\times$ 10$^{10}$ & $BVRizJK_s$3.6$\mu$m4.5$\mu$m \\
3114 & 1 $\times$ 10$^{11}$ & 2 $\times$ 10$^{10}$ & $VRizJK_s$3.6$\mu$m4.5$\mu$m \\
3132 & 1 $\times$ 10$^{11}$ & 4 $\times$ 10$^{10}$ & $BVRizJK_s$3.6$\mu$m4.5$\mu$m \\
3214 & 6 $\times$ 10$^{10}$ & 2 $\times$ 10$^{10}$ & $BVRiz$3.6$\mu$m4.5$\mu$m \\
3424 & 1 $\times$ 10$^{10}$ & 7 $\times$ 10$^{09}$ & $BVRizJK_s$3.6$\mu$m4.5$\mu$m \\
3767 & 1 $\times$ 10$^{10}$ & 3 $\times$ 10$^{09}$ & $BVRizJK_s$3.6$\mu$m4.5$\mu$m \\
3860 & 3 $\times$ 10$^{11}$ & 1 $\times$ 10$^{11}$ & $VRizJK_s$3.6$\mu$m4.5$\mu$m \\
4159 & 3 $\times$ 10$^{11}$ & 3 $\times$ 10$^{10}$ & $BVRizJK_s$ \\
4199 & 3 $\times$ 10$^{11}$ & 3 $\times$ 10$^{10}$ & $BVRizJK_s$ \\
4235 & 9 $\times$ 10$^{10}$ & 9 $\times$ 10$^{09}$ & $BVRizJK_s$3.6$\mu$m4.5$\mu$m \\
4325 & 2 $\times$ 10$^{11}$ & 4 $\times$ 10$^{10}$ & $BVRizJK_s$3.6$\mu$m4.5$\mu$m \\
4479 & 4 $\times$ 10$^{11}$ & 8 $\times$ 10$^{10}$ & $BVRizJK_s$3.6$\mu$m4.5$\mu$m \\
4503 & 1 $\times$ 10$^{11}$ & 2 $\times$ 10$^{10}$ & $BVRizJK_s$3.6$\mu$m4.5$\mu$m \\
4525 & 3 $\times$ 10$^{10}$ & 6 $\times$ 10$^{09}$ & $BVRizJK_s$3.6$\mu$m4.5$\mu$m \\
4553 & 9  $\times$ 10$^{10}$ & 2 $\times$ 10$^{10}$ & $BVRizJK_s$ \\
5137 & 7 $\times$ 10$^{10}$ & 1 $\times$ 10$^{10}$ & $BVRizJK_s$3.6$\mu$m4.5$\mu$m \\
5221 & 2 $\times$ 10$^{11}$ & 3 $\times$ 10$^{10}$ & $BVRizJK_s$ \\
5222 & 2 $\times$ 10$^{11}$ & 4 $\times$ 10$^{10}$ & $BVRizJK_s$ \\
5397 & 1 $\times$ 10$^{11}$ & 1 $\times$ 10$^{10}$ & $BVRizJK_s$3.6$\mu$m4.5$\mu$m \\
5801 & 4 $\times$ 10$^{11}$ & 4 $\times$ 10$^{10}$ & $BVRizJK_s$3.6$\mu$m4.5$\mu$m \\
5997 & 4 $\times$ 10$^{11}$ & 8 $\times$ 10$^{10}$ & $BVRiz$3.6$\mu$m4.5$\mu$m \\
6052 & 4 $\times$ 10$^{10}$ & 2 $\times$ 10$^{10}$ & $BVRizJK_s$3.6$\mu$m4.5$\mu$m \\
6064 & 5 $\times$ 10$^{10}$ & 4 $\times$ 10$^{10}$ & $BVRizJK_s$3.6$\mu$m4.5$\mu$m \\
6115 & 3 $\times$ 10$^{10}$ & 1 $\times$ 10$^{10}$ & $BVRizJK_s$3.6$\mu$m4.5$\mu$m \\
6383 & 3 $\times$ 10$^{10}$ & 3 $\times$ 10$^{09}$ & $BVRizJK_s$3.6$\mu$m4.5$\mu$m \\
6392 & 1 $\times$ 10$^{11}$ & 2 $\times$ 10$^{10}$ & $VRizJK_s$ \\
6592 & 8 $\times$ 10$^{10}$ & 3 $\times$ 10$^{10}$ & $BVRiz$ \\
6673 & 4 $\times$ 10$^{10}$ & 4 $\times$ 10$^{10}$ & $BVRiz$ \\
6688 & 1 $\times$ 10$^{10}$ & 1 $\times$ 10$^{10}$ & $BVRiz$ \\
7546 & 6 $\times$ 10$^{08}$ & 1 $\times$ 10$^{09}$ & $BVRiz$ \\
7581 & 1 $\times$ 10$^{10}$ & 4 $\times$ 10$^{09}$ & $BVRizJK_s$ \\
8193 & 2 $\times$ 10$^{10}$ & 1 $\times$ 10$^{10}$ & $BVRizJK_s$3.6$\mu$m4.5$\mu$m \\
8529 & 4 $\times$ 10$^{09}$ & 4 $\times$ 10$^{08}$ & $BVRizJK_s$ \\
8686 & 2 $\times$ 10$^{11}$ & 3 $\times$ 10$^{10}$ & $BVRizJK_s$3.6$\mu$m4.5$\mu$m \\
8709 & 3 $\times$ 10$^{11}$ & 5 $\times$ 10$^{10}$ & $BVRizJK_s$3.6$\mu$m4.5$\mu$m \\
9067 & 2 $\times$ 10$^{09}$ & 2 $\times$ 10$^{09}$ & $BVRizJK_s$3.6$\mu$m4.5$\mu$m \\
9197 & 1 $\times$ 10$^{10}$ & 1 $\times$ 10$^{10}$ & $BVRiz$ \\
9204 & 9 $\times$ 10$^{09}$ & 2 $\times$ 10$^{09}$ & $BVRizJK_s$ \\
9227 & 1 $\times$ 10$^{11}$ & 2 $\times$ 10$^{10}$ & $BVRizJK_s$3.6$\mu$m4.5$\mu$m \\
9354 & 1 $\times$ 10$^{11}$ & 2 $\times$ 10$^{10}$ & $BVRizJK_s$3.6$\mu$m4.5$\mu$m \\
9465 & 9 $\times$ 10$^{10}$ & 2 $\times$ 10$^{10}$ & $BVRizJK_s$3.6$\mu$m4.5$\mu$m \\
9466 & 1 $\times$ 10$^{11}$ & 1 $\times$ 10$^{10}$ & $BVRizJK_s$3.6$\mu$m4.5$\mu$m \\
9470 & 3 $\times$ 10$^{10}$ & 1 $\times$ 10$^{10}$ & $BVRizJK_s$3.6$\mu$m4.5$\mu$m \\
30145 & 5 $\times$ 10$^{09}$ & 9 $\times$ 10$^{08}$ & $BVRizJK_s$3.6$\mu$m4.5$\mu$m \\
30337 & 4 $\times$ 10$^{09}$ & 7 $\times$ 10$^{09}$ & $BVRizJK_s$3.6$\mu$m4.5$\mu$m \\
30917 & 8 $\times$ 10$^{10}$ & 4 $\times$ 10$^{10}$ & $BVRizJK_s$3.6$\mu$m4.5$\mu$m \\
40950 & 2 $\times$ 10$^{10}$ & 5 $\times$ 10$^{09}$ & $BVRizJK_s$3.6$\mu$m4.5$\mu$m \\
\hline  
\end{tabular}
\end{table}

We defined environmental density regions using the weak-lensing maps of Jee et al.~(\cite{jee05}, \cite{jee11}) and the definitions used in Demarco et al.~(\cite{dem10}) for J0152.7.  Demarco et al.~(\cite{dem10}) used weak-lensing maps to determine dark-matter densities defined as fractions of the critical density for J0152.7.  We avoided using distance from the cluster center as a measure of the environment, since the center is not necessarily well-defined for a merging, non-virialized cluster such as J0152.7.  Also, smaller haloes detectable in dark-matter density could have a similar effect on galaxy evolution to the regions just outside a cluster core.  The $\kappa$ values are defined as a ratio of the dark-matter density in a given location ($\Sigma$) to the critical density ($\Sigma_{c}$) for that galaxy cluster, as described in Eq.~2 of Jee et al.~(\cite{jee11}).  Demarco et al.~(\cite{dem10}) defined the high-density regions of J0152.7 as densities of over 20 $\sigma_{DM}$, where $\sigma_{DM}$ = 0.0057 $\times$ $\Sigma_{c}$ (i.e. 1 $\sigma_{DM}$ corresponds to $\kappa$ = 0.0057), and $\Sigma_{c}$ = 3650 M$_{\odot}$ /pc$^2$ (Blakeslee et al.~\cite{bla06}).  The contours representing these regions are shown in Fig.~3, with the thick dashed black lines representing intermediate-density-region boundaries and the thick solid black lines representing high-density-region boundaries.

The high-density regions in J0152.7 roughly correspond to everything within at least three positive-$\kappa$ contours ($\kappa$ $>$ 0.12) on the Jee et al.~(\cite{jee05}) weak-lensing map of J0152.7.  The intermediate-density regions of J0152.7 (5 $\sigma_{DM}$ to 20 $\sigma_{DM}$) approximately correspond to anything within at least one but fewer than three positive-$\kappa$ contours (0.04 $<$ $\kappa$ $<$ 0.12).  For J1252.9, we defined density regions comparable to those of J0152.7 by using similar $\kappa$ values to make the cutoffs.  Since the contour scale of the J1252.9 weak-lensing map is very similar to that of J0152.7, we could easily create density regions similar to those of J0152.7.  The J1252.9 high-density regions were within three or more Jee et al.~(\cite{jee11}) contours ($\kappa$ $>$ 0.111).  The intermediate-density regions were within at least one but fewer than three contours (0.037 $<$ $\kappa$ $<$ 0.111).  The J1252.9 low-density regions were outside any of the positive-$\kappa$ contours ($\kappa$ $<$ 0.037).  Some intermediate-density regions in J0152.7 appear to extend beyond the edges of the mass map, and one J1252.9 member lies just beyond the density map in what is likely still part of the intermediate-density region (see Fig.~2).  We consider this galaxy part of the intermediate-density regions, although counting it as a member of the low-density regions will not significantly affect our results.

We note that due to the mass-sheet degeneracy, the absolute scales of the weak-lensing maps are not well-determined.  However, the most important aspect for our density-region classifications is the relative difference between regions of high and low density.

Stellar-mass subsamples were chosen in order to ensure a substantial population of galaxies in all mass subsamples of J1252.9 for comparison with J0152.7.  Stellar masses for J1252.9 galaxies were calculated using the technique of Gobat et al.~(\cite{gob08}).  Each galaxy had at least five photometric bands ranging from $B$ to {\it{Spitzer}} 4.5$\mu$m.  For J1252.9 galaxies in Tab.~1 possessing only optical photometric data, reddening-free fits were used so as not to overestimate reddening and thus overestimate the stellar mass.  The value obtained by these fits is expected to be accurate within about a factor of 2.  For galaxies with infrared photometry ($J+K_s$ and/or {\it{Spitzer}}), the code was allowed to estimate the internal reddening of the galaxy.  For passive galaxies, the reddening never exceeded A$_v$ = 0.5 or E$(B-V)$ = 0.2, as was also found in Rettura et al.~(\cite{ret06}).  

Stellar-mass estimates for J1252 galaxies, along with uncertainties and bands used for the fit, are listed in Tab.~2.  Most uncertainties were estimated from the spectrophotometric fitting code using a variety of redshifts between z = 1.22 and z = 1.25, including the measured redshift of the galaxy.  We fit with a variety of redshifts to estimate errors because any difference in redshift within this range is more likely to reflect the galaxy's peculiar velocity than its relative distance along our line of sight compared to other cluster members.  If the uncertainty determined from the code is less than 10\% for galaxies with IR photometry ($J$, $K_s$, 3.6$\mu$m, or 4.5$\mu$m) or less than 100\% (a factor of two) for galaxies without these bands, we set the uncertainty value to 10\% or 100\% of the calculated stellar mass, respectively.

After determining the stellar masses, we found that about half of the J1252.9 sample, including three passive galaxies, had masses less than 7 $\times$ 10$^{10}$ M$_{\odot}$.  We thus chose this stellar-mass limit to divide the high-mass vs. low-mass samples for both clusters.  The J0152.7 stellar masses are mostly those used in Demarco et al.~(\cite{dem05,dem10}).  A correlation between these stellar masses and VLT High Acuity Wide-field $K$-band Imager (HAWK-I) $K$-band magnitudes (Lidman et al.~\cite{lid12}, discussed below) was used to estimate stellar mass for 20 galaxies in J0152.7 with no New Technology Telescope (NTT) Son of ISAAC (SofI) $K$-band data.

\subsection{Spectroscopic completeness and stacking weights}

\begin{figure*}
\centering
\includegraphics[width=18cm]{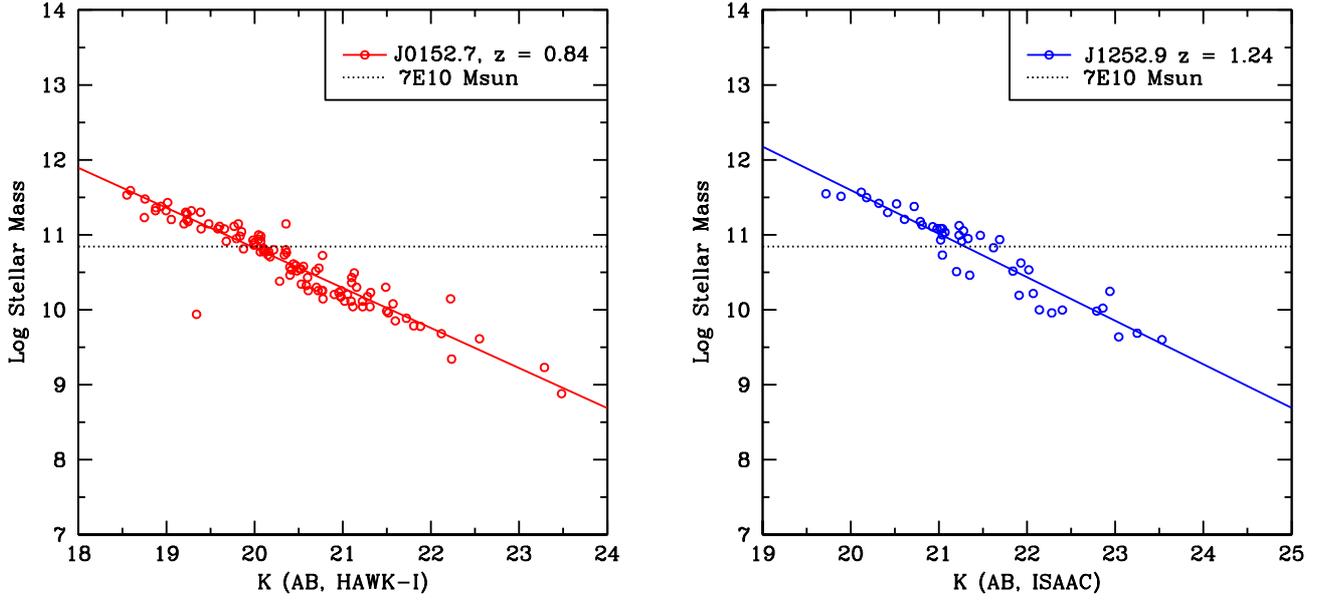}
\caption{Stellar mass as a function of K-band magnitude for spectroscopic cluster members of J0152.7 (left) and J1252.9 (right). The horizontal dotted line in each panel represents 7 $\times$ 10$^{10}$ M$_{\odot}$, the dividing line between high-mass and low-mass samples.}
\label{FigStelMassProxy}
\end{figure*}

\begin{figure*}
\centering
\includegraphics[width=18cm]{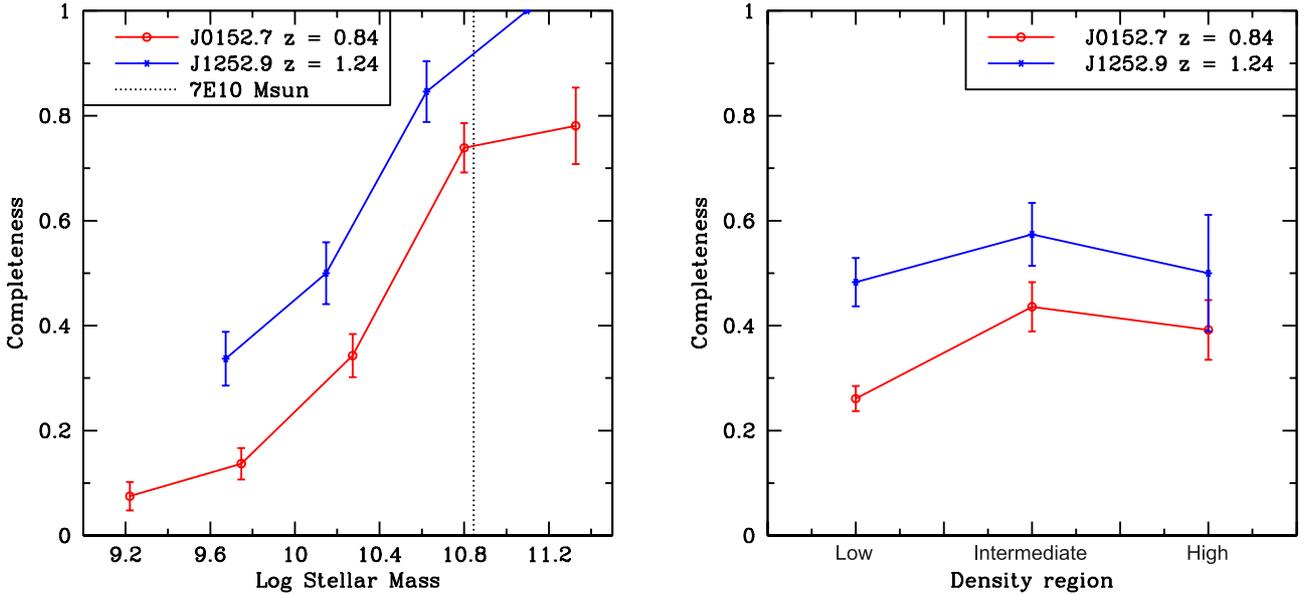}
\caption{Spectroscopic completeness fractions as a function of stellar mass (left) and density region (right) for J0152.7 (red) and J1252.9 (blue).  The horizontal dotted line in the left panel represents 7 $\times$ 10$^{10}$ M$_{\odot}$, the dividing line between high-mass and low-mass samples. Completeness fractions for both clusters show a notable relationship to stellar mass but only a weak relationship to density.  Uncertainties are determined using the bootstrap method resampled 10000 times.}
\label{FigComp}
\end{figure*}

Weighting spectra before stacking allows for improvement in the S/N of the spectra (and thus more accurate index measurements) and correction for spectroscopic completeness. In Demarco et al.~(\cite{dem10}), stacked spectra were weighted according to S/N in the continuum windows near the H$\delta$ line, without concerns for spectroscopic completeness within the subsamples.  In Muzzin et al.~(\cite{muz12}), completeness corrections for stellar mass in environment-based subsamples and for environment in stellar-mass-based subsamples were applied before averaging index values.  In order to determine how to assign weights to the spectra, we estimated stellar-mass and environmental-density completeness functions in a fashion similar to Muzzin et al.~(\cite{muz12}) as described below to determine and apply appropriate incompleteness corrections for both clusters.

In order to construct the stellar-mass completeness functions, we first chose a photometric proxy for stellar mass that could be used to estimate the stellar masses of non-members and unconfirmed candidates.  In Muzzin et al.~(\cite{muz12}), this stellar mass proxy was the  {\it{Spitzer}}) 3.6$\mu$m band, which was used to identify the clusters in the infrared-selected survey.  For our analysis, we chose $K$-band magnitudes as a stellar mass proxy.  The $K$-band had the closest relationship to the calculated stellar masses for J0152.7 galaxies, and $K_s$ was the most commonly available non-blended infrared photometric band for J1252.9.  

The $K$-band magnitudes used as a stellar mass proxy in J1252.9 were the VLT/ISAAC $K_s$ AB aperture magnitudes from Rettura et al.~(\cite{ret10}).  These same $K$-band magnitudes had been used as part of the Gobat et al.~(\cite{gob08}) method to calculate stellar masses of confirmed J1252.9 members when available.  For J0152.7, we used MAG\_AUTO Vega magnitudes determined from the HAWK-I $K$-band imaging (Lidman et al.~\cite{lid12}) with Source Extractor version 2.8.6 (Bertin \& Arnouts \cite{ber96}).  A Vega calibration for these data had been performed by Lidman et al.~(\cite{lid12}), and the Vega magnitudes were converted to AB by adding 1.85.  The HAWK-I data were used instead of SofI data because the former covered the full angular area of the cluster, while the latter missed about 20 members.  We used a 3$\sigma$ detection limit in order to include as many faint galaxies as possible in the catalog.  MAG\_AUTO was used for J0152.7 because in the previously existing SofI data (Demarco et al.~\cite{dem10}), $K$-band MAG\_AUTO had the strongest correlation to calculated stellar mass.  

For both clusters, the ranges of available magnitudes and colors for confirmed cluster members were used to select photometric member candidates.  Members with calculated stellar masses were then used to determine a correlation between $K$-band magnitude and stellar mass.  This correlation was then used to estimate the stellar mass of all photometric member candidates in the sample, including the confirmed members, to construct the completeness functions.  

The stellar mass vs. $K$ relations for spectroscopically-confirmed members are shown in Figure 4.  The linear fits have a standard deviation of 0.190 dex for J1252.9 and 0.137 dex for J0152.7.  This indicates that $K$-band magnitude is accurate as a stellar mass proxy within a factor of about 1.5--1.6 for J1252.9 and within a factor of about 1.4 for J0152.7.  The slopes of these relations differ by about 9\% ($-$0.534 for J0152.7 vs. $-$0.581 for J1252.9), suggesting a similar relationship between $K$-band magnitude and stellar mass for both clusters.   

After estimating the stellar masses via $K$-band magnitude, photometric member candidates were organized into an appropriate number of bins according to their estimated stellar mass (four bins for J1252.9 and five bins for J0152.7).  Completeness in each bin was calculated according to Eq.~1 of Muzzin et al.~(\cite{muz12}) after counting the confirmed members, confirmed nonmembers (field galaxies), and total candidates in each bin.

To determine environmental density completeness, we used the same subsamples of photometric member candidates defined above.  We used the modified map in Fig.~4 of Demarco et al.~(\cite{dem10}) and the map in Fig.~23 of Jee et al.~(\cite{jee11}) to visually determine the locations of candidates with respect to the high, intermediate, and low-density regions.  Figures 2 and 3 of this paper also show the density-region boundaries we used.  Once all objects were assigned to a density region, we used Eq.~1 of Muzzin et al.~(\cite{muz12}) to assign a completeness value to each of the three density regions for each of the two clusters.   

The completeness functions in mass and density for both clusters are shown in Fig.~5.  Uncertainties are determined using the bootstrap method (random resampling of original data, in this case lists of members, confirmed nonmembers, and unconfirmed candidates) resampled 10000 times.  The raw J1252.9 completeness values are higher than those of J0152.7.  This may be because J1252.9 had fewer member candidates (especially at moderate to high stellar mass or $K$-band flux) in the angular area from which most targets were selected.  The completeness as a function of density is nearly flat (especially in J1252.9), and the completeness as a function of stellar mass/$K$-band flux has a similar shape for both clusters.  The relative completeness fractions compared to the highest-stellar-mass bins and most-populated density bins are therefore similar in both clusters.  The overall density completeness values are under 60\%, since all density regions have similarly large numbers of faint, spectroscopically-unconfirmed galaxies.

The completeness functions for both clusters are of similar shape, but there is an offset between the two clusters, and a slight difference in shape most notable in the density completeness functions.  The offset is not important in terms of combining samples, since we do not combine samples from two different clusters.  However, our mass-based samples include only two categories, above and below 7 $\times$ 10$^{10}$ M$_{\odot}$, both of which span multiple mass-completeness bins in both clusters. We therefore correct for mass completeness in all subsamples, and also density completeness in mass-based subsamples, before stacking the spectra.  The total stacking weight is thus the product of the S/N around 4100 {\AA}, the inverse of the stellar-mass completeness, and, for mass-based subsamples, the inverse of the density completeness.  (Density-based subsamples span only one completeness bin in density and therefore need only mass-based completeness correction.)  Mass-based completeness is assigned by bin, i.e., all objects in the same cluster that occupy the same stellar mass bin used in calculating the completeness function are assigned the same completeness.  In each cluster, two galaxies lie below the mass cutoff of the lowest-mass completeness bin for their cluster and are thus excluded from stacking.  We note that there is no major qualitative difference in our results when we stack the spectra with or without completeness corrections.  Star-forming fractions are estimated in Sect.~4 both with and without the relevant completeness corrections for mass and density.

\subsection{Definition of H6a index}

The influence of a young stellar population in a galaxy spectrum is traditionally measured by an H$\delta$ index, such as H$\delta_A$ (Worthey \& Ottaviani \cite{wor97}).  At high redshifts, however, H$\delta$ can have low S/N or even be redshifted out of the usable part of a ground-based optical/near-infrared spectrum.  In the case of FORS2, where the usable portion of the spectrum extends to about 10000 {\AA}, the red continuum of H$\delta_A$ lies beyond the useful part of the spectrum at z = 1.403.  

A possible alternative to this index is the H6 equivalent width as defined by Demarco et al.~(\cite{dem10}).  It is measured in the same way as other equivalent widths (see e.g. Worthey \& Ottaviani \cite{wor97}), with flux measured and compared in blue (B) and red (R) pseudo-continuum bands and a central band (I) measuring the absorption or emission feature:

\begin{equation}
EW = (\lambda_2-\lambda_1)[1-2F_I/(F_B+F_R)]
\end{equation}

In the above equation, $\lambda_2$ and $\lambda_1$ are the upper and lower bounds, respectively, of the wavelength range defining the central band, I; $F_I$ is the total flux in the central band; and $F_B$ and $F_R$ are the total fluxes in the blue and red pseudo-continuum bands, respectively.  In the original H6 index, the central band I spans 3868-3908 \AA, the blue pseudo-continuum B spans 3852-3864 \AA, and the red pseudo-continuum R spans 3911-3921 \AA.

As seen in Fig.~6 using Bruzual \& Charlot (\cite{bc03}) (BC03) models, the H6 equivalent width reaches a minimum after the peak and increases at older ages.   The turn-over occurs at an age of about 2.5 Gyr for a solar-metallicity (Z$_{\odot}$) simple stellar population (SSP), the solid blue curve in Fig.~6.  For a 2.5 Z$_{\odot}$ SSP (solid red curve in Figure 5), the turn-over occurs shortly after 1 Gyr.  This is due to the increasing influence of blue CN-band absorption at the H6 wavelength and the blue pseudo-continuum, while the red pseudo-continuum is nearly unaffected.  Therefore, if an H6-based spectral index is to be used as a reliable tracer of young stellar populations like H$\delta$, it must account for the growing influence of the CN band at older ages.

We have thus defined a new H6 equivalent width, H6a, with bandwidths adjusted such that the index values stabilize instead of steadily and perceptibly rising at old ages.  The central bandwidth (I) was reduced from 40 {\AA} to 20 {\AA}, spanning the range 3878.5-3898.5 {\AA}.  This narrower central band allows for adjustment of pseudo-continuum bands to include CN absorption, so that the flux in all bands is depressed by CN in similar amounts.  The new red pseudo-continuum (R) ranges from 3900.5 to 3910.5 {\AA}, to avoid the areas least affected by CN and therefore keep the continuum and central bands at more comparable levels.  The new blue pseudo-continuum (B) ranges from 3846-3876 {\AA} in order to cover a large portion of blue CN band absorption.  The wide blue band can also help recover some of the S/N lost by narrowing the central band.

\begin{figure}
\centering
\includegraphics[width=8.5cm]{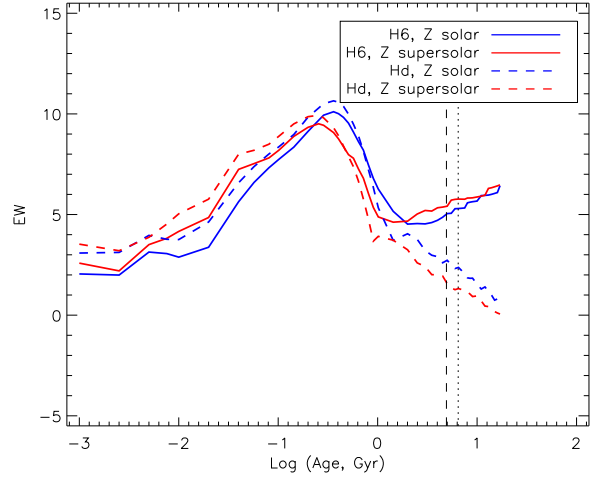}
\caption{H$\delta$ (dashed lines) and H6 (solid lines) equivalent widths as a function of stellar-population age, for BC03 SSP models smoothed to 9 {\AA} resolution.  Note that the H6 equivalent width begins to rise at intermediate ages due to the influence of the CN band, while H$\delta$ continues to fall with age after a peak around 300 Myr.  The vertical black dashed and dotted lines represent the ages of the universe at the redshifts of J1252.9 and J0152.7, respectively.}
\label{FigHdEv}
\end{figure}

\begin{figure}
\centering
\includegraphics[width=8.5cm]{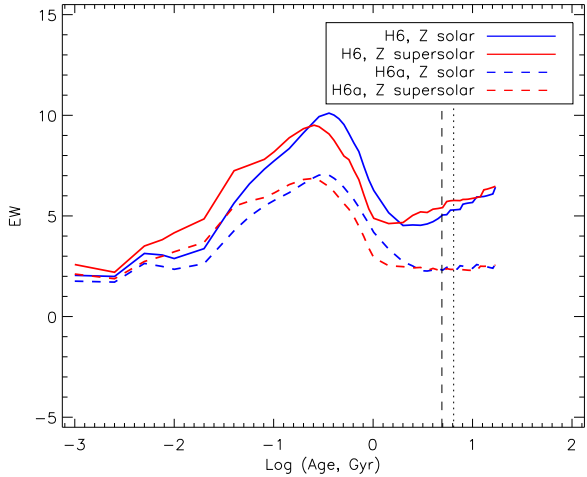}
\caption{Evolution of the H6a index (dashed lines) and the Demarco et al.~(\cite{dem10}) H6 index (solid lines) for BC03 models at 9 {\AA} resolution.  Unlike the older H6 index, H6a does not increase perceptibly with increasing age beyond 2.5 Gyr for Z$_{\odot}$ models or 1 Gyr for 2.5 Z$_{\odot}$ models.}
\label{FigH6Evol}
\end{figure}

\begin{table*}
\caption{Star-forming Fractions}
\label{tab3}
\centering
\begin{tabular}{llccccccccc}
\hline\hline  
Cluster & Sample & Ngal & SF Frac., wtd. & $\sigma$ & SF Frac., raw & $\sigma$  \\
 & &  &  &   \\ 
\hline
J1252.9 & HDMD & 8 & 0.27 & 0.16 & 0.25 & 0.15 \\
J1252.9 & IDMD & 19 & 0.53 & 0.12 & 0.42 & 0.11  \\
J1252.9 & LDMD & 23 & 0.82 & 0.07 & 0.74 & 0.09  \\
J1252.9 & High-mass & 26 & 0.24 & 0.09 & 0.23 & 0.08  \\
J1252.9 & Low-mass & 24 & 0.91 & 0.05 & 0.88 & 0.07  \\
J1252.9 & Total & 50 & 0.66 & 0.07 & 0.54 & 0.07 \\
J0152.7 & HDMD & 22 & 0.00 & 0.00 & 0.00 & 0.00  \\
J0152.7 & IDMD & 37  & 0.24 & 0.09 & 0.22 & 0.07  \\
J0152.7 & LDMD & 65 & 0.70 & 0.07& 0.52 & 0.06  \\
J0152.7 & High-mass & 41 & 0.18 & 0.06 & 0.17 & 0.06  \\
J0152.7 & Low-mass & 83 & 0.62 & 0.07 & 0.42 & 0.05  \\
J0152.7 & Total & 124 & 0.57 & 0.06 & 0.34 & 0.04   \\
\hline
\end{tabular}
\tablefoot{HDMD = high dark-matter density, IDMD = intermediate dark-matter density, LDMD = low dark-matter density.  As specified in Sect.~3.1, high-mass and low-mass refer to above and below 7 $\times$ 10$^{10}$ M$_{\odot}$, respectively.  Uncertainties in star-forming fractions were calculated using the bootstrap method resampled 10000 times.  Completeness weights for the weighted (completeness-corrected) star-forming fractions in Col.~4 were the inverse of the mass completeness for density-based samples, and the inverse of the product of mass and density completeness for all other samples.  The star-forming fractions in Col.~5 are without completeness correction. }
\end{table*}

The behavior of the H6a index with age compared to the original H6 index is shown in Fig.~7.   While the original H6 index climbs steadily at older ages, H6a becomes nearly flat beyond the turn-over point.  The H6a index is slightly more sensitive to spectral resolution and S/N than the original H6 index due to the narrower central bandwidth.  However, as we will see below, the greater accuracy of the H6a index compared to the original H6 index in measuring pure H6 line absorption (as opposed to CN) can make up for the higher uncertainties due to these factors.  Neither index shows strong sensitivity to reddening across the wavelength range of the band definitions.

If H6a is used instead of H$\delta$ to identify k+a and a+k (post-starburst) spectral features, one can make the appropriate distinctions by comparing the values of H6a and H$\delta$ at ages at which 3 {\AA} $<$ EW(H$\delta$) $<$ 8 {\AA} for k+a and EW(H$\delta$) $>$ 8 {\AA} for a+k.  At 9 {\AA} resolution for a solar-metallicity BC03 model, EW(H$\delta$) $=$ 3 {\AA} corresponds to EW(H6a) $=$ 2.26 {\AA}.  An EW(H$\delta$) $=$ 8 {\AA} corresponds to EW(H6a) $=$ 5.37 {\AA} at this same resolution and metallicity.  

When tested on a sample of eight galaxies in J0152.7 identified by Demarco et al.~(\cite{dem05}) as post-starburst k+a or a+k galaxies, seven of the eight galaxies were identified as having EW(H6a) above the threshold of 2.26 {\AA}.  All eight galaxies had higher EW(H6a) than the J0152.7 high-density passive stack (see Sect.~4), which was below this threshold.  

The Demarco et al.~(\cite{dem10}) H6 index also found the same seven galaxies to be above the threshold corresponding to ages at which H$\delta$ $>$ 3 {\AA} (EW(H6) $>$ 4.59 {\AA}).  However, the J0152.7 high-density passive stack also had an EW(H6) above this value, and also above the EW(H6) of the one post-starburst galaxy which H6 failed to identify.  This result demonstrates the greater reliability of H6a compared to the older H6 index for distinguishing between young and old stellar populations.  The statistical significance of the H6 differences between post-starburst galaxies and the old passive population was greater with H6a than with the older H6 index for four of the seven H6 post-starburst galaxies. Therefore, H6a shows better performance than the Demarco et al.~(\cite{dem10}) H6 index in distinguishing between old passive galaxy spectra and k+a spectra.  This makes H6a a good candidate to identify post-starburst galaxies from ground-based optical spectroscopy at higher redshifts than is currently feasible with H$\delta$.  The red pseudo-continuum of H6a does not reach the FORS2 useful limit of 10000  {\AA} until z = 1.557, vs. z = 1.403 for H$\delta_A$.

\section{Results}

Table 3 lists the star-forming fractions of the environment- and stellar-mass-based samples of each cluster, with and without completeness weighting.   Table 4 lists the various galaxy cluster sub-populations and the spectral indices from their stacked spectra: H6a, $D_n$4000 (Kauffmann et al.~\cite{kau03}), and [OII] equivalent width (Tran et al.~\cite{tra03}) and their uncertainties.

In our analysis, we do not attempt to convert [OII] equivalent widths or fluxes into star formation rates.  The varying shapes, sizes, and orientations of the galaxies in the two clusters make correction of [OII] flux for slit losses difficult.  The lack of high-quality mid- or far-IR data also makes correcting for reddening difficult.  However, we can still use [OII] to determine star-forming fractions, and compare [OII] equivalent widths among star-forming and combined samples.

\subsection{Mixed galaxy subsamples}

\begin{figure*}
\centering
\includegraphics[width=18cm]{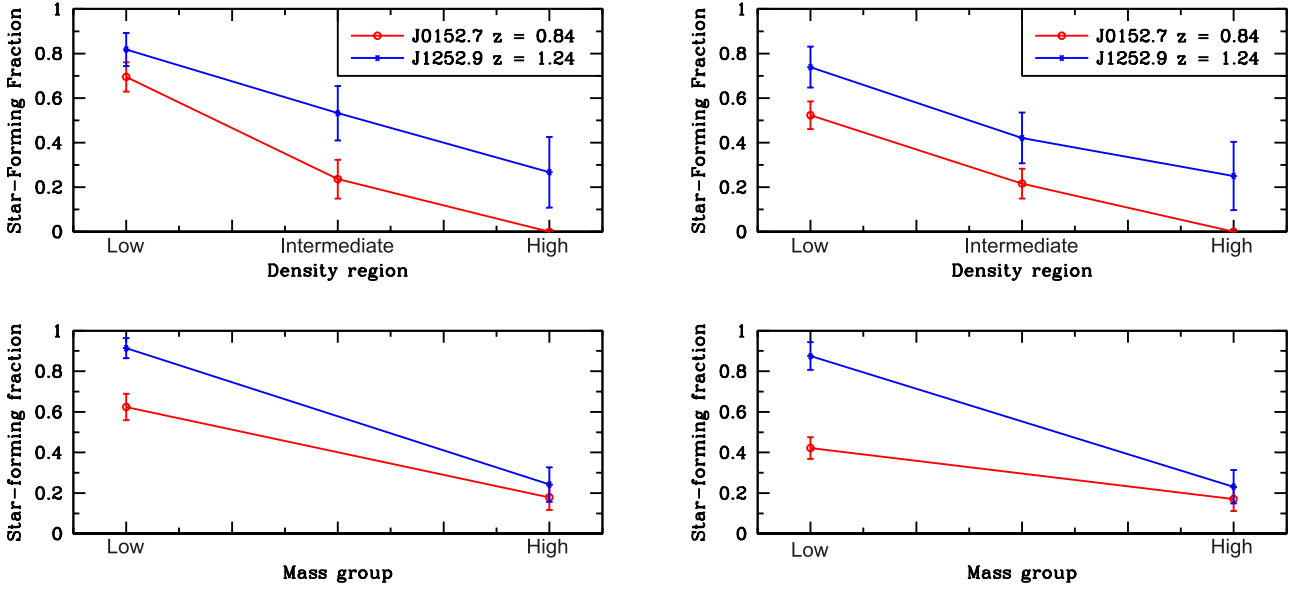}
\caption{Star-forming fractions for J1252.9 (blue) and J0152.7 (red) as a function of environmental density (top) and stellar mass (bottom).  The star-forming fraction of the low-mass sample of J1252.9 is significantly higher than that of J0152.7, and at least marginally higher in all but the high-mass sample.}
\label{FigH6def}
\end{figure*}

\begin{figure*}
\centering
\includegraphics[width=18cm]{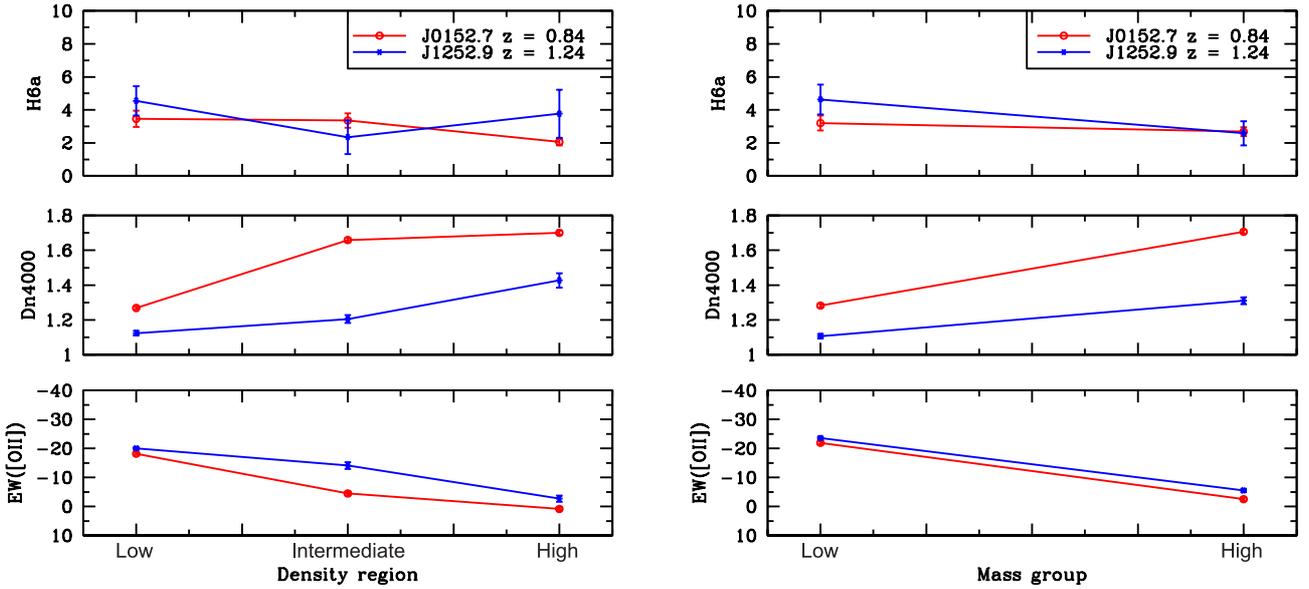}
\caption{Spectral index values for passive and star-forming galaxies combined as a function of environmental density and stellar mass for J0152.7 (red) and J1252.9 (blue).  Trends are generally similar with both density and stellar mass for both clusters, although in $D_n$4000 the changes are greatest between low and intermediate density in J0152.7 and greatest between intermediate and high density in J1252.9.}
\label{FigIndDiff}
\end{figure*}

\begin{figure*}
\centering
\includegraphics[width=18cm]{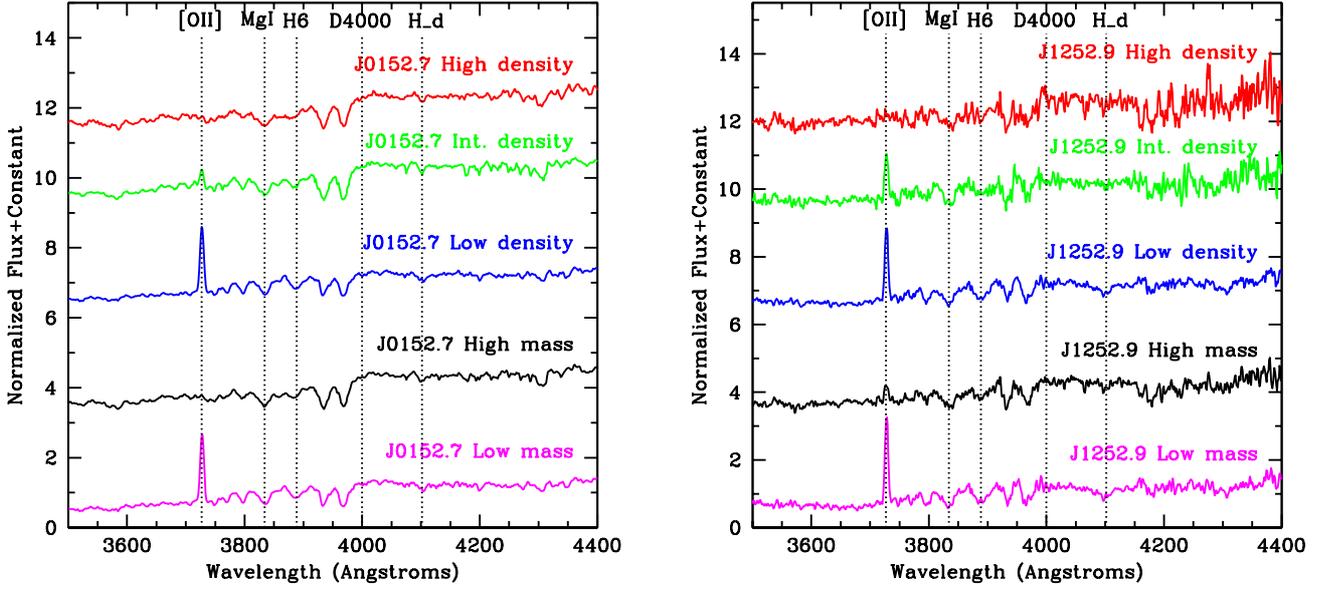}
\caption{Stacked spectra, including both passive and star-forming galaxies, for the density and stellar-mass groupings of J0152.7 (left) and J1252.9 (right).  Spectra of similar density and stellar mass in the two clusters show similar features in general, though with stronger [OII] and weaker 4000 {\AA} breaks visible in J1252.9.  All wavelengths shown are in the rest frame.}
\label{FigStackTotal}
\end{figure*}

\begin{table*}
\caption{Spectral Indices}
\label{tab4}
\centering
\begin{tabular}{llccccccc}
\hline\hline  
Cluster & Sample & Ngal & H6a & $\sigma$ & $D_n$4000 & $\sigma$ & [OII] & $\sigma$ \\
 & &  & {\AA} & {\AA} &  &  & {\AA} & {\AA} \\ 
\hline
J1252 & HDMD & 8 & 3.8 & 1.5 & 1.428 & 0.041 & -2.7 & 1.1 \\
J1252 & IDMD & 19 & 2.3 & 1.0 & 1.205 & 0.023 & -14.1 & 1.2 \\
J1252 & LDMD & 23 & 4.5 & 0.9 & 1.124 & 0.013 & -20.1 & 0.5 \\
J1252 & High-mass & 26 & 2.6 & 0.7 & 1.311 & 0.020 & -5.6 & 0.6 \\
J1252 & Low-mass & 24 & 4.6 & 0.9 & 1.106 & 0.013 & -23.6 & 0.5 \\
J1252 & HDMD Passive & 6 & 4.2 & 1.4 & 1.504 & 0.057 & -0.8 & 1.2 \\
J1252 & IDMD Passive & 11 & 2.5 & 1.7 & 1.405 & 0.041 & -1.2 & 1.7 \\
J1252 & IDMD Star-forming & 8 & 4.1 & 1.6 & 1.116 & 0.016 & -33.3 & 1.2 \\
J1252 & LDMD Passive & 6 & 3.5 & 1.5 & 1.357 & 0.032 & -6.4 & 1.2 \\
J1252 & LDMD Star-forming & 17 & 4.8 & 1.0 & 1.084 & 0.015 & -26.1 & 0.5 \\
J1252 & High-mass Passive & 20 & 2.2 & 0.8 & 1.428 & 0.029 & -1.1 & 0.8 \\
J1252 & High-mass Star-forming & 6 & 5.4 & 1.6 & 1.114 & 0.016 & -13.0 & 1.2 \\
J1252 & Low-mass Passive & 3 & 9.1 & 2.5 & 1.259 & 0.082 & -8.0 & 2.6 \\
J1252 & Low-mass Star-forming & 21 & 4.4 & 1.0 & 1.078 & 0.015 & -28.3 & 0.5 \\
J0152 & HDMD & 22 & 2.1 & 0.2 & 1.701 & 0.012 & 0.8 & 0.5 \\
J0152 & IDMD & 37 & 3.4 & 0.4 & 1.659 & 0.012 & -4.5 & 0.6 \\
J0152 & LDMD & 65 & 3.5 & 0.5 & 1.269 & 0.008 & -18.2 & 0.4 \\
J0152 & High-mass & 41 & 2.7 & 0.3 & 1.707 & 0.011 & -2.6 & 0.5 \\
J0152 & Low-mass & 83 & 3.2 & 0.5 & 1.282 & 0.010 & -21.9 & 0.5 \\
J0152 & HDMD Passive & 22 & 2.1 & 0.2 & 1.701 & 0.012 & 0.8 & 0.5 \\
J0152 & IDMD Passive & 29 & 3.5 & 0.3 & 1.707 & 0.012 & -1.1 & 0.7 \\
J0152 & IDMD Star-forming & 8 & 2.5 & 0.8 & 1.497 & 0.031 & -17.3 & 1.0 \\
J0152 & LDMD Passive & 31 & 2.9 & 0.3 & 1.585 & 0.012 & -1.7 & 0.5 \\
J0152 & LDMD Star-forming & 34 & 3.9 & 0.6 & 1.120 & 0.010 & -33.7 & 0.5 \\
J0152 & High-mass Passive & 34 & 2.6 & 0.2 & 1.726 & 0.009 & 0.3 & 0.4 \\
J0152 & High-mass Star-forming & 7 & 2.4 & 0.7 & 1.545 & 0.019 & -10.9 & 0.8 \\
J0152 & Low-mass Passive & 48 & 2.5 & 0.3 & 1.578 & 0.013 & -1.2 & 0.6 \\
J0152 & Low-mass Star-forming & 35 & 3.8 & 0.6 & 1.096 & 0.011 & -38.5 & 0.5 \\
\hline
\end{tabular}
\tablefoot{Mass and density categories in Col.~2 are defined as follows: HDMD = high dark-matter density, IDMD = intermediate dark-matter density, LDMD = low dark-matter density, High-mass is above 7 $\times$ 10$^{10}$ M$_{\odot}$, and low-mass is below 7 $\times$ 10$^{10}$ M$_{\odot}$.  Passive galaxies lack a visual [OII] detection, while star-forming galaxies show definite or highly probable [OII].}
\end{table*}

The most notable difference between J1252.9 and J0152.7 is in their star-forming fractions, given for the clusters as a whole and as a function of environment and stellar mass in Tab.~3.  The star-forming fractions as a function of environment and stellar mass are also shown in Fig.~8, both corrected and uncorrected for completeness.  Uncertainties in star-forming fractions were determined using the bootstrap method resampled 10000 times per original sample.  Star-forming fractions in J1252.9 are higher than in J0152.7 overall and in all samples except for the high-mass sample, although these differences are marginal except in the low-mass sample.  Both clusters show similar variation of star-forming fraction as a function of environmental density, and J1252.9 shows a more dramatic difference than J0152.7 with respect to stellar mass.  The drop in star-forming fraction in both clusters with both denser environment and higher stellar mass is consistent with the results of Sobral et al.~(\cite{sob11}) at z $\sim$ 0.84 using H$\alpha$.

The star-forming fraction of J1252.9 galaxies in the low-mass sample, which includes galaxies of up to 7 $\times$ 10$^{10}$ M$_{\odot}$, is 91.4\% (corrected for completeness).  This value is about 4$\sigma$ higher than the star-forming fraction among J0152.7 galaxies in the same stellar mass range, 62.4\% (corrected for completeness).  

Figure 9 shows the spectral indices as a function of environment and stellar mass for the two clusters.  Figure 10 displays the stacked spectra themselves.  Over all, the trends with density region and stellar mass are similar for both clusters.   $D_n$4000 increases for both clusters as a function of environmental density and stellar mass, while [OII] decreases.  Both trends are consistent with the differences in star-forming fraction seen in Fig.~8.  The differences in $D_n$4000 and in [OII] are similar to what was found for combined samples of passive and star-forming galaxies in Muzzin et al.~(\cite{muz12}).  In all density and stellar mass groupings, $D_n$4000 is higher in J0152.7, as might be expected for its lower redshift.  In J0152.7, H6a shows a modest increase (about 3.6 $\sigma$) in the low- and intermediate-density regions compared to the high-density regions.  

In Fig.~9, $D_n$4000 changes most dramatically in J0152.7 between the low- and intermediate-density regions, while in J1252.9 the biggest change in $D_n$4000 is seen between the intermediate- and high-density regions.  This difference can also be seen in Fig.~10.  In Sect.~5, we will discuss the implications of the various differences found between the mixed passive and star-forming galaxy subsamples in J0152.7 and J1252.9.

\subsection{Passive galaxy subsamples}

\begin{figure*}
\centering
\includegraphics[width=18cm]{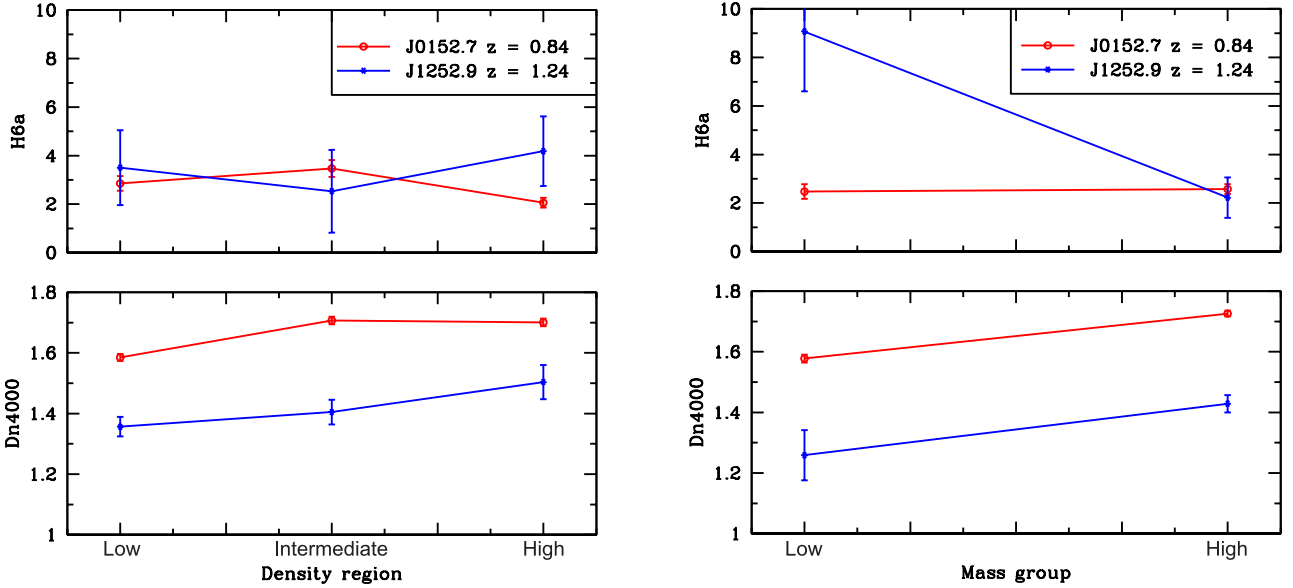}
\caption{Spectral index values as a function of environmental density and stellar mass for passive galaxies in J0152.7 (red) and J1252.9 (blue).  Differences in $D_n$4000 with stellar mass and density are similar in both clusters, apart from a slight but notable decrement in $D_n$4000 for J0152.7 passive galaxies in the low-density regions with respect to their counterparts in higher-density regions.  There is also a marginal (due to low S/N) elevation of H6a among J1252.9 low-mass galaxies with respect to J1252.9 high-mass galaxies.  A modestly significant H6a difference is found between the J0152.7 intermediate-density population and the J0152.7 high-density population.}
\label{FigIndDiffPa}
\end{figure*}

\begin{figure*}
\centering
\includegraphics[width=18cm]{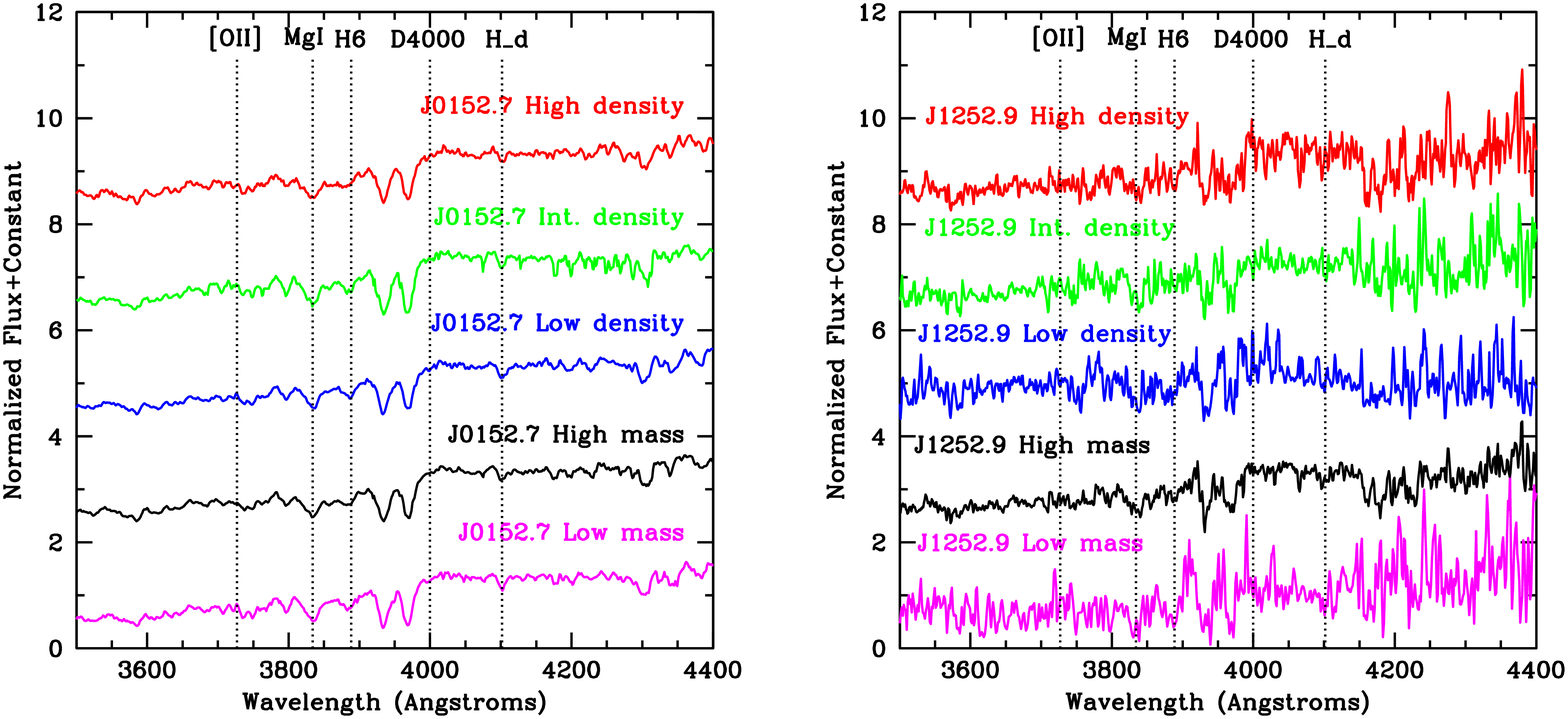}
\caption{Stacked spectra of passive galaxies for the density and mass groupings of J0152.7 (left) and J1252.9 (right).  Spectra of similar density and stellar mass in the two clusters show similar features in general, though with weaker 4000 {\AA} breaks in J1252.9, and the strong (but noisy) H6 line visible in the J1252.9 low-mass sample.  All wavelengths shown are in the rest frame.}
\label{FigStackPa}
\end{figure*}

The H6a and $D_n$4000 indices for passive galaxies in each cluster as a function of density and stellar mass are shown in Fig.~11.  The stacked spectra of these samples are shown in Fig.~12.  The first notable trend is that the J0152.7 passive samples all have higher $D_n$4000 values than those of J1252.9.  The statistical significance of these differences is above 3$\sigma$ in all samples. 

In the H6a vs.~stellar mass plot of Fig.~11 and the stacked spectra of Fig.~12, the J1252.9 low-mass passive sample appears to have a strong, though noisy, H6 feature.  However, the poor S/N makes this result inconclusive, as it corresponds to only a difference in H6a strength of 2.7$\sigma$ from the J1252.9 high-mass passive subsample.  A small but substantial (4.1$\sigma$) H6a enhancement is seen in the intermediate-density subsample of J0152.7 compared to the high-density subsample of the same cluster.  This modest H6 enhancement is also visible in Fig.~12.  Due to the lower S/N of the J1252.9 passive galaxy samples, no H6a enhancements of at least 3$\sigma$ are observed in J1252.9.

\subsection{Star-forming galaxy subsamples}

\begin{figure*}
\centering
\includegraphics[width=18cm]{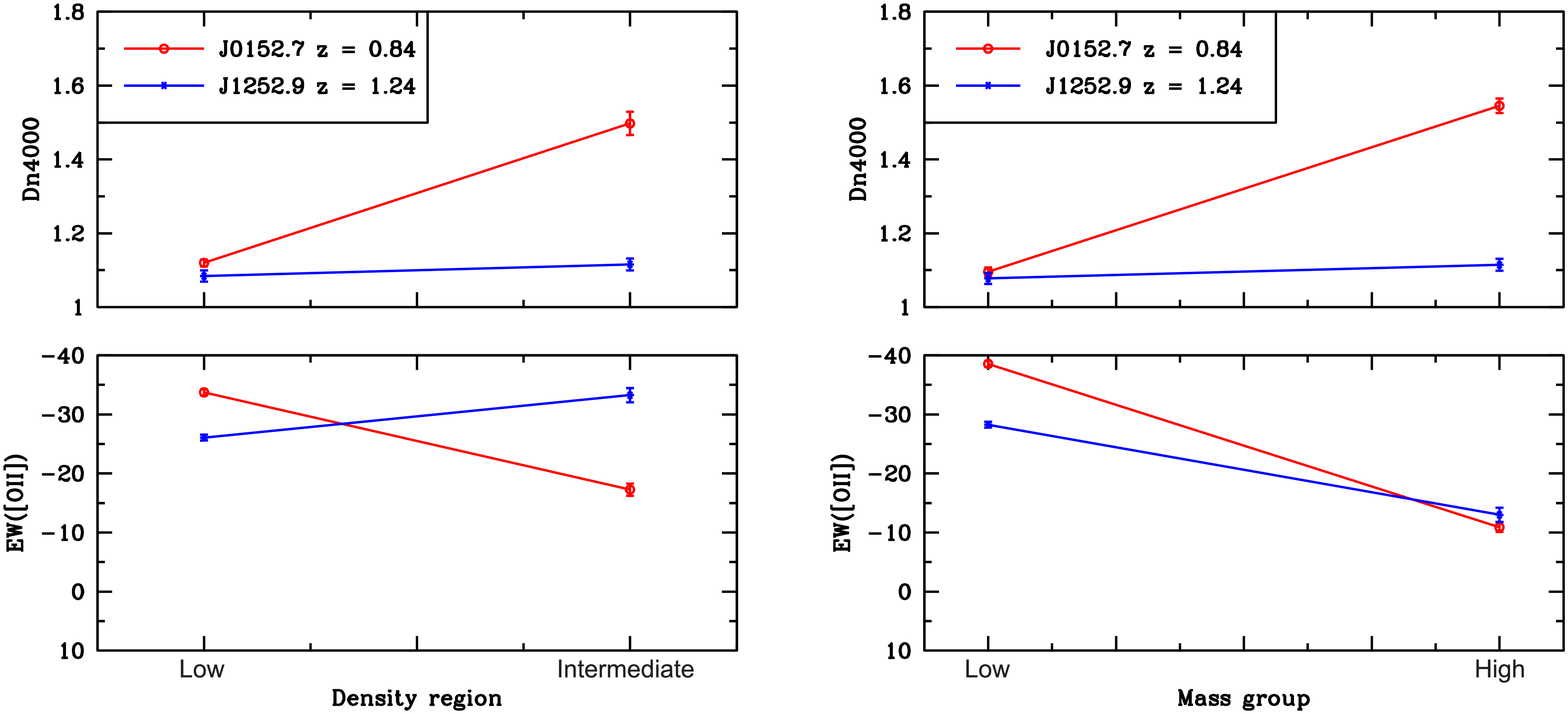}
\caption{Spectral index values as a function of environmental density and stellar mass for star-forming galaxies in J0152.7 (red) and J1252.9 (blue).  The $D_n$4000 values depend strongly on environmental density and stellar mass in J0152.7, but not in J1252.9.  EW([OII]) decreases similarly with stellar mass in both clusters.}
\label{FigIndDiffSf}
\end{figure*}

\begin{figure*}
\centering
\includegraphics[width=18cm]{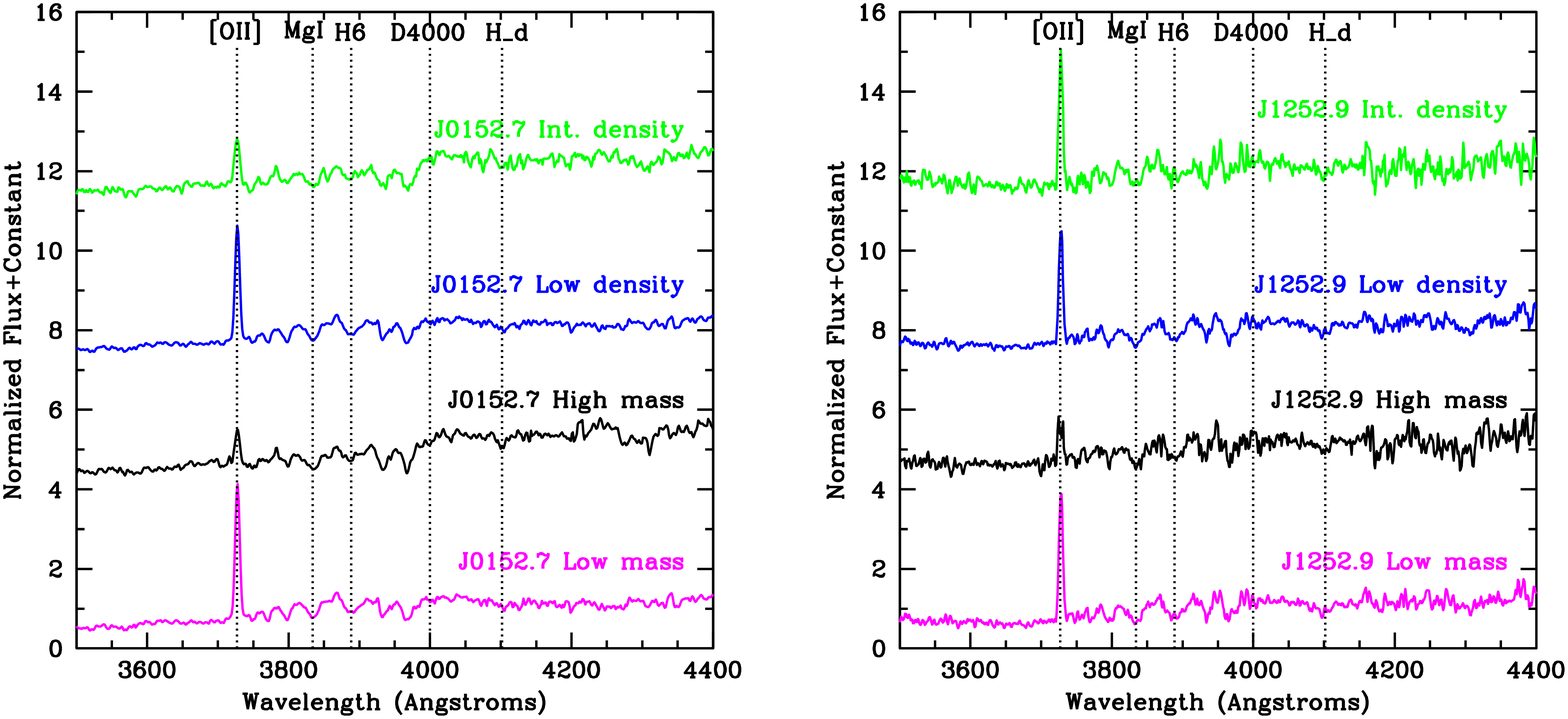}
\caption{Stacked spectra of star-forming galaxies for the density and mass groupings of J0152.7 (left) and J1252.9 (right).  In J0152.7, the 4000 {\AA} break is substantially stronger in the intermediate-density and high-mass samples than in other samples from either cluster.  All wavelengths shown are in the rest frame.}
\label{FigStackSf}
\end{figure*}

The $D_n$4000 and [OII] spectral indices of the star-forming galaxy populations of the two clusters are compared in Fig.~13.  H6a is not shown, since it is only useful to distinguish between young (up to 1 Gyr) and old ($>$ 2.5 Gyr) stellar populations and not between two young or two old stellar populations.  In particular, a young stellar population with relatively low H6a could either be only a few Myr old or 1 Gyr old.  In Fig.~14, stacked star-forming spectra are shown for both clusters.  In J1252.9, little change with stellar mass or density is seen except for a notable decrease of EW([OII]) with increasing stellar mass, consistent with Muzzin et al.~(\cite{muz12}).  In fact, [OII] appears slightly enhanced in the intermediate-density regions of J1252.9 compared to the low-density regions, although this result disappears if mass completeness is not taken into account.  

On the other hand, $D_n$4000 varies substantially in the J0152.7 star-forming galaxies with both environmental density and stellar mass.  The $D_n$4000 values of high-mass and intermediate-density J0152.7 star-forming galaxies are higher than those of even high-mass passive galaxies in J1252.9. They are also within 3$\sigma$ of the $D_n$4000 values for low-mass and low-density J0152.7 passive galaxies.

There is also an EW([OII]) decrement at intermediate density in J0152.7.  In the completeness-corrected samples shown in Figs.~13 and 14, this is quite substantial (15$\sigma$), although if completeness is not accounted for, this difference drops to about 5$\sigma$.  Since we do not have detailed information on dust absorption or slit losses for galaxies in either cluster, we cannot be certain how much of this slight [OII] difference is related to specific star-formation rates or extinction.

\section{Discussion}

In Sect.~4.1 we found that star-forming fractions were higher in J1252.9 than in J0152.7 in all environments, and substantially higher (4$\sigma$) among galaxies less massive than 7 $\times$ 10$^{10}$ M$_{\odot}$. We note that the lowest-mass passive galaxy in J1252.9 is above the Demarco et al.~(\cite{dem10}) mass cutoff for intermediate-mass red sequence galaxies. This implies that all of the J1252.9 counterparts of the Demarco et al.~(\cite{dem10}) low-mass red sequence of J0152.7, galaxies less massive than 2.7 $\times$ 10$^{10}$ M$_{\odot}$, are still forming stars.  There is thus a substantial difference visible between the two clusters, in which many more galaxies of intermediate to low stellar mass have stopped forming stars at z=0.84 as compared to z=1.24.

The low number of spectroscopically confirmed low-mass passive galaxies in J1252.9 is consistent with previous studies (e.g., Tanaka et al.~\cite{tan07}, Gilbank et al.~\cite{gil08}) finding a lack of faint, red, quiescent galaxies at high redshifts.   Also, Demarco et al.~(\cite{dem10}) found that faint blue red-sequence members and low-mass red-sequence members probably stopped forming their stars at z$\sim$1, about 800 Myr after the epoch of J1252.9.

If the trends in star-forming fractions with environment, stellar mass, and redshift found in Sect.~4.1 are extended to higher redshifts, we would eventually expect to find star formation even in massive cluster core galaxies.  Hayashi et al.~(\cite{hay10}) found nearly equal [OII]-based star-forming fractions in the outskirts and core of the z = 1.46 galaxy cluster XMMXCS 1215 (610 Myr before the epoch of J1252.9).  In a follow-up paper (Hayashi et al.~\cite{hay11}), they found that the cluster-core [OII]-emitting galaxies were red and had multi-band colors and line ratios suggestive of AGN contribution.

Further evidence for substantial star formation among bright galaxies in cluster cores has recently been found in clusters and protoclusters at z$\sim$1.6 (940 Myr before the epoch of J1252.9).  In the Fassbender et al.~(\cite{fas11}) cluster at z=1.56, the third-brightest galaxy in the core shows starburst activity.  In the Muzzin et al.~(\cite{muz13}) cluster at z=1.63, nine out of 12 confirmed members have [OII] emission, including the brightest confirmed member.   And in the Papovich et al.~(\cite{pap10}) cluster at z=1.62, the averaged (stacked) spectrum of the seven closest spectroscopically confirmed galaxies to the cluster core shows substantial [OII] emission.

We also found in Sect.~4.1 that in J0152.7, the $D_n$4000 values were more similar between the high- and intermediate-density regions, while in J1252.9 they were more similar between the intermediate- and low-density regions. This result suggests that the relative evolutionary states of the average cluster galaxy in the intermediate-density regions of J1252.9 and J0152.7 are different.  In J1252.9, the galaxies in intermediate-density regions still largely resemble the galaxies of the low-density outskirts in terms of $D_n$4000 and absorption lines.  Their [OII] emission, however, is similar to that of the J0152.7 intermediate-density galaxies.  In J0152.7, the galaxies of the intermediate-density region are beginning to look more like the galaxies of the cluster core in most respects.

In Sect.~4.2, we found substantial offsets at all stellar masses and environments between the $D_n$4000 values of passive galaxies in J0152.7 and J1252.9. The corresponding differences in typical stellar population age, based on the BC03 models, range from 1.8--3.0 Gyr for a Z$_{\odot}$ model and 0.6--1.3 Gyr for a 2.5 Z$_{\odot}$ model.  The $D_n$4000 differences would thus be consistent with 1.5 Gyr passive evolution at metallicities intermediate between these two models.  Also notable is the fact that in J0152.7, the low-density and low-mass $D_n$4000 values are substantially lower (more than 9$\sigma$) than those of other J0152.7 samples.  This suggests a higher fraction of intermediate-age stars in these galaxies.

We also found a slight enhancement of H6a, of about 4$\sigma$, in passive galaxies in the intermediate-density regions of J0152.7 compared to the high-density regions of the same cluster.  This result may represent higher populations of relatively young, post-starburst passive galaxies in the intermediate- and low-density regions compared to the high-density regions. 

In Sect.~4.3, we found similar variation of [OII] with stellar mass both clusters.  This result is expected if high-mass galaxies have lower specific star formation rates than low-mass galaxies, as found in Muzzin et al.~(\cite{muz12}).  $D_n$4000, however, behaved quite differently between the two clusters, being highly enhanced (and similar to the values for J1252.9 passive galaxies) in the intermediate-density and high-mass subsamples of J0152.7.  The $D_n$4000 index is moderately sensitive to reddening due to its relatively broad wavelength coverage, so part of the difference between J0152.7 and J1252.9 could be due to reddening.  However, it would take a very high reddening value, $E(B-V)$ $\sim$ 2.0 with a Calzetti et al.~(\cite{cal00}) reddening law, to increase $D_n$4000 from 1.1 to 1.55.  Within J1252.9, intermediate-density star-forming galaxies were only slightly more reddened than low-density star-forming galaxies ($\delta$A$_V$ $\sim$ 0.1) according to our composite stellar population fits.  Lower-mass J1252.9 galaxies with reddening estimates were in fact more reddened than higher-mass galaxies, albeit by a modest amount ($\delta$A$_V$ $\sim$ 0.4). Therefore, we doubt that the high $D_n$4000 values of intermediate-density and high-mass galaxies in J0152.7 can be attributed to systematically high reddening.

While Muzzin et al.~(\cite{muz12}) found $D_n$4000 of star-forming galaxies to be unrelated to cluster-centric radius in all stellar mass bins, we found a substantial environmental effect on this index in J0152.7 but not in J1252.9.  Although the reasons for this are unclear, part of this effect may be due to the fact that four of the eight star-forming galaxies of the J0152.7 intermediate-density regions are of high stellar mass.  In contrast, only one of the seven J1252.9 intermediate-density star-forming galaxies is of high stellar mass.  However, even high-mass star-forming galaxies in J1252.9 do not have higher $D_n$4000 than their low-mass counterparts.  Perhaps these galaxies were more recently accreted in J1252.9 than in J0152.7.  Intermediate-density star-forming galaxies in J0152.7 may be among the next to join the passive red sequence, with their stellar populations already beginning to resemble those of relatively young passive galaxies.

Over all, galaxy populations in the z=0.84 cluster J0152.7 are more mature, i.e., less likely to be star-forming and with older underlying stellar populations, than in the z=1.24 cluster J1252.9.  The former implies that significant star-formation quenching has occurred in the 1.5 Gyr separating the two clusters.  This could be due to a variety of factors: self-quenching mechanisms in star-forming galaxies simply having more time to work, group pre-processing, or if the cluster-crossing timescales are on the order of the time elapsed between the two cluster redshifts, cluster crossings.  Ram pressure, harassment, and strangulation could quench star formation as a galaxy crosses the cluster, and then the galaxy could return to its original distance from the cluster center (but a different location) in the outskirts.  Wetzel et al.~(\cite{wet13a}) make a strong case suggesting that low-mass galaxies in cluster outskirts could be quenched by falling into the cluster and then returning to the outskirts, commonly referred to as backsplash.  In their simulations, star-forming galaxies that have crossed the cluster core can continue to experience diminished star formation even as they return to the cluster outskirts.  Low-mass galaxies within 1--2.5 virial radii had an especially high likelihood (up to 40\%) of being backsplash galaxies compared to those at greater cluster-centric distances.  These backsplash galaxies could also still be found up to several virial radii from a group or cluster center, creating enhanced fractions of passive or low-star-formation-rate galaxies compared to the field.  Galaxies that had crossed a group or cluster core would often not show substantial quenching until 2--4 Gyr after initial infall (Wetzel et al.~\cite{wet13b}).  

It is possible that a number of J1252.9 galaxies in the intermediate- to low-density regions are just falling in to the cluster, while many of their J0152.7 counterparts have already crossed.  Such an effect might explain the rapidly maturing star-forming populations in the intermediate-density regions of J0152.7, as well as the overall lower star-forming fractions, especially among low-mass galaxies. Tanaka et al.~(\cite{tan07}) and Muzzin et al.~(\cite{muz12}), among others, suggest greater susceptibility to environmental effects among low-mass galaxies at z$\sim$1.  

The crossing time from the virial radius of J0152.7 (1.4 Mpc) at a de-projected velocity dispersion of 1300 km s$^{-1}$ (Girardi et al.~\cite{gir05}; Demarco et al.~\cite{dem10}) is about 1.08 Gyr using the estimate of Sarazin (\cite{sar88}).  The crossing time at a typical intermediate-density-region distance of 600 kpc is about 0.46 Gyr.  This would allow for easy passage of these galaxies through the cluster core and back to the intermediate-density regions in less than 1.5 Gyr.

Our crossing-time estimates suggest that most galaxies in the intermediate- to low-density regions of J0152.7 have had time to pass through the densest regions in the 1.5 Gyr since the epoch of J1252.9.   Star-forming galaxies in the intermediate-density regions of J0152.7 would thus be more likely to start to show quenching effects than their J1252.9 counterparts.  Low-mass galaxies outside the core might also be more likely to be passive than their J1252.9 counterparts as a result of more cluster crossings.  However, there are also differences in $D_n$4000 seen among high-mass star-forming galaxies in J0152.7 and J1252.9.  The stellar-mass effect may be primarily related to self-quenching mechanisms (e.g. supernova feedback) having had more time to work in  J0152.7.

\section{Summary}

We have compared stacked spectra of spectroscopically confirmed members of the galaxy clusters J1252.9 and J0152.7, separated by 1.5 Gyr of cosmic time, sorted according to environment and stellar mass.  Further sorting according to passive or star-forming status was also performed to look for evolution with redshift, stellar mass, and environment among these sub-populations.  Our analysis included 20 new members of J1252.9 confirmed with FORS2.  We also defined a new spectral index, H6a, which measures the strength of the Balmer H6 line without significant increases due to CN in old stellar populations.

Comparing the full galaxy populations in the two clusters sorted according to environmental density and stellar mass, we note that the star-forming fraction among J1252.9 galaxies below 7 $\times$ 10$^{10}$ M$_{\odot}$ is substantially higher than that of J0152.7.  The difference is about 4 $\sigma$, suggesting a modest but notable reduction in star formation among low-mass galaxies within 1.5 Gyr.  The variation of star-forming fraction with environmental density is similar in both clusters, and the star-forming fractions in J1252.9 are higher in all environments.

Among passive galaxies, variation in $D_n$4000 was generally similar with both density and stellar mass for both clusters, and the total $D_n$4000 values were higher for all samples in J0152.7 than in J1252.9.  This difference in the indices was consistent with 1.5 Gyr of passive evolution at modestly super-solar metallicities.

\begin{acknowledgements}
J. N. acknowledges the support provided by FONDECYT postdoctoral research grant \# 3120233.  R.D. gratefully acknowledges the support provided by
the BASAL Center for Astrophysics and Associated Technologies (CATA), and by FONDECYT grant N. 1130528.  C.L. is the recipient of an Australian Research Council Future Fellowship (program number FT0992259).
\end{acknowledgements}

\end{document}